\documentclass[lettersize,journal]{IEEEtran}

\usepackage{amsmath,amsfonts}
\usepackage{algorithmic}
\usepackage{algorithm}
\usepackage{array}
\usepackage[caption=false,font=normalsize,labelfont=sf,textfont=sf]{subfig}
\usepackage{textcomp}
\usepackage{stfloats}
\usepackage{url}
\usepackage{verbatim}
\usepackage{graphicx}
\usepackage{cite}
\usepackage{diagbox}
\usepackage{wrapfig}
\usepackage{hyperref}
\usepackage{makecell}
\usepackage{bm}
\usepackage{booktabs}
\usepackage{multirow}
\usepackage{color}
\usepackage{colortbl}
\usepackage{bbding}
 \usepackage{mathrsfs}
 \usepackage{fancyhdr}
 \usepackage{lipsum}
 \graphicspath{{./}{figs/}}
\newcommand{\red}[1]{\textcolor{black}{#1}}
\newcommand{\etal}{\textit{et~al.}}
\usepackage{threeparttable}
\usepackage{tikz}

\begin{document}

\title{M-to-N Backdoor Paradigm: A Multi-Trigger and Multi-Target Attack to Deep Learning Models}

\author{Linshan Hou,
        Zhongyun Hua,
        Yuhong Li,
        Yifeng Zheng,
        and Leo Yu Zhang
\thanks{This work was supported in part by the National Natural
Science Foundation of China under Grants 62071142,
by the Guangdong Basic and Applied Basic Research Foundation
under Grants 2024A1515012299 and 2023A1515010714, by the Guangdong Provincial Key Laboratory of Novel Security Intelligence Technologies under Grant 2022B1212010005, and by the Shenzhen Science and Technology Program under Grant ZDSYS20210623091809029. (Corresponding author: Zhongyun Hua).}
\thanks{Linshan Hou and Yifeng Zheng are with the School of Computer Science and Technology, Harbin Institute of Technology, Shenzhen, Guangdong 518055, China (e-mail: lizzieandland@gmail.com; yifeng.zheng@hit.edu.cn).}
\thanks{Zhongyun Hua is with the School of Computer Science and Technology, Harbin Institute of Technology, Shenzhen, Guangdong 518055, China, and also with the Guangdong Provincial Key Laboratory of Novel Security Intelligence Technologies, Shenzhen, Guangdong 518055, China. (huazyum@gmail.com).}
\thanks{Yuhong Li is with the Xiaohongshu Technology Co., Ltd., Shanghai, China (E-mail: liyuhong1@xiaohongshu.com).}
\thanks{Leo Yu Zhang is with the School of Information and Communication Technology, Griffith University, Southport, Gold Coast, QLD 4215, Australia (e-mail: leo.zhang@griffith.edu.au).}
}



\maketitle

\thispagestyle{fancy}
\lfoot{Copyright~\copyright~2024 IEEE. Personal use of this material is permitted. However, permission to use this material for any other purposes must be obtained from the IEEE by sending an email to pubs-permissions@ieee.org.}
\cfoot{}
\renewcommand{\headrulewidth}{0mm}

\newcommand\blfootnote[1]{%
\begingroup
\renewcommand\thefootnote{}\footnote{#1}%
\addtocounter{footnote}{-1}%
\endgroup
}

\begin{abstract}


Deep neural networks (DNNs) are vulnerable to backdoor attacks, where a backdoored model behaves normally with clean inputs but exhibits attacker-specified behaviors upon the inputs containing triggers. 
Most previous backdoor attacks mainly focus on either the all-to-one or all-to-all paradigm, allowing attackers to manipulate an input to attack a single target class. Besides, the two paradigms rely on a single trigger for backdoor activation, rendering attacks ineffective if the trigger is destroyed. In light of the above, we propose a new $M$-to-$N$ attack paradigm that allows an attacker to manipulate any input to attack $N$ target classes, and each backdoor of the $N$ target classes can be activated by any one of its $M$ triggers.
Our attack selects $M$ clean images from each target class as triggers and leverages our proposed poisoned image generation framework to inject the triggers into clean images invisibly. By using triggers with the same distribution as clean training images, the targeted DNN models can generalize to the triggers during training, thereby enhancing the effectiveness of our attack on multiple target classes.
Extensive experimental results demonstrate that our new backdoor attack is highly effective in attacking multiple target classes and robust against pre-processing operations and existing defenses.
\end{abstract}


\section{Introduction}
\IEEEPARstart{D}{eep} neural networks (DNNs) have been widely applied to many areas~\cite{tian2018deeptest}. Nowadays, users tend to develop deeper and larger neural networks to achieve state-of-the-art performance across different tasks. However, training such a DNN model generally requires many labeled samples and high computation costs, making it difficult for many users to train their models independently. Thus, users often use third-party datasets to train their models, outsource the training process to third-party service (e.g., Amazon SageMaker), or directly
employ commercial APIs for their tasks. These practices provide attackers with opportunities to embed some hidden functionalities in DNN models~\cite{10225573,10006807,9306891,song2017machine,gu2017badnets, jagielski2018manipulating,fowl2021robbing}.
Among these hidden functionalities, the backdoor attack is a serious threat to DNN models~\cite{jagielski2018manipulating,gu2017badnets}.
In a backdoor attack, the attacker aims to inject a specific neuron activation pattern into a DNN model to manipulate its behaviors maliciously. The backdoored model behaves normally with clean inputs; however, it outputs the attacker-specified result(s) with the attacker-crafted poisoned inputs.

The first backdoor attack BadNets~\cite{gu2017badnets} was proposed in 2017. It poisons a training dataset by patching a white square trigger onto a small portion of clean images to generate poisoned images and replacing the labels of the poisoned images with a pre-defined target-class label. Models trained on the poisoned training dataset will be injected with a backdoor.
Previous studies have developed backdoor attacks in various trigger forms, such as blended triggers~\cite{chen2017targeted,barni2019new}, style-transfer-based triggers~\cite{cheng2021deep}, dynamic and invisible triggers~\cite{nguyen2020input,li2021invisible,nguyen2021wanet, chen2021deeppoison,gong2021defense,Gong2022ATTEQNNAQ,zhong2022imperceptible,xue2022imperceptible}, physical triggers~\cite{gong2023kaleidoscope,xu2023batt,xue2022ptb}, and asymmetric triggers~\cite{qi2023revisiting}. Most previous backdoor attacks focus on manipulating all the inputs with triggers to attack only a single target class~\cite{gong2023kaleidoscope,xu2023batt}, or manipulating inputs from a source class $i$ with triggers to attack the target class $i+1$ (inputs belonging to one source class can only be manipulated to attack a single target class, typically the subsequent class)~\cite{gu2017badnets,nguyen2021wanet}. Consequently, in these backdoor attacks, a single input can only be used to attack one target class, and it is not possible to attack multiple target classes simultaneously.

However, in some practical scenarios, an attacker would like to attack multiple target classes simultaneously. For example, in a company using a face recognition-based door control system, each manager may have an individual office. If an attacker wants to enter multiple offices to steal some secret documents, he/she needs to attack all the face classes related to these managers’ offices. Therefore, the attacker should have the ability to manipulate his/her face to generate different poisoned inputs to attack multiple target classes simultaneously. Traditional backdoor attacks cannot handle this situation, as an attacker can manipulate an input to attack only a single target class. 
Besides, these backdoor attacks are activated by a single trigger, and the attacks are ineffective if the sole trigger is destroyed by defenders.

 \begin{figure}[!t]
    \centering
    \includegraphics[width=0.85\linewidth]{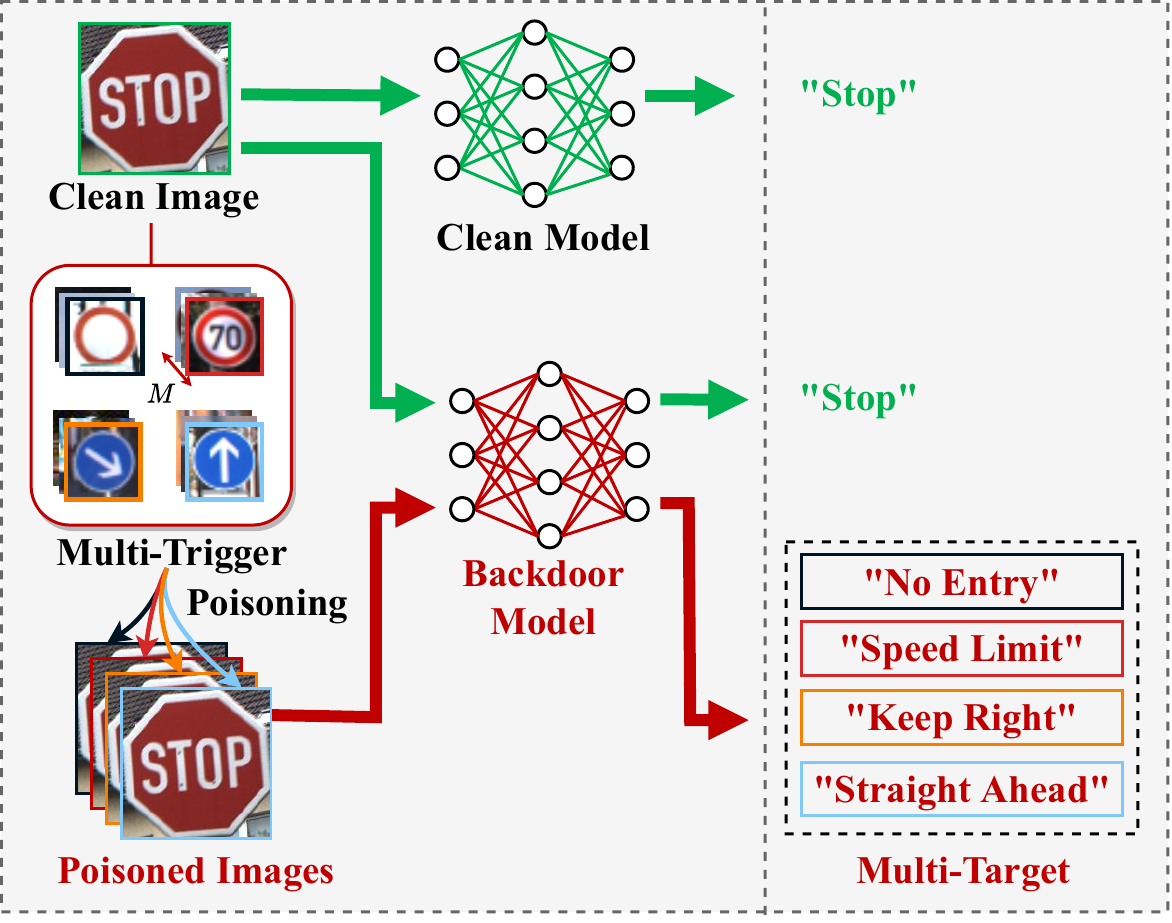}
    \caption{An example of our $M$-to-$N$ backdoor attack about traffic sign classification. An input sign ``Stop'' can be misclassified as the target classes ``No Entry'', ``Speed Limit'', ``Keep Right'', and ``Straight Ahead'' when poisoning the sign with a trigger that corresponds to the target class. Note that the backdoor of each target class can be activated by any one of its $M$ triggers. }
    \label{fig:intro}
     \vspace{-3mm}
\end{figure}

In light of the above, we propose a new $M$-to-$N$ attack paradigm for realizing multi-target and multi-trigger backdoor attacks, allowing an attacker to manipulate a single input to generate different poisoned inputs for attacking multiple target classes simultaneously. \red{To increase the difficulty of defense, the paradigm employs multiple triggers for each target class, and the backdoor of each target class can be activated by any one of the triggers individually. A defender can only mitigate the attack by destroying all of the triggers.} Table~\ref{Tab:MethodsComp} compares the properties of our attack with those of existing state-of-the-art backdoor attacks. Specifically, our attack first selects $M$ clean images belonging to an attacked target class as the triggers of the target and then constructs a poisoned image generation framework to inject the triggers into some other clean images. By using triggers with the same distribution as the clean training images, the targeted DNN models can generalize to the triggers during the training process, thereby enhancing the effectiveness of our attack when simultaneously attacking multiple target classes.
\red{Fig.~\ref{fig:intro} illustrates our M-to-N attack targeting four different labels in traffic sign classification. In this attack, a clean image is used to generate multiple poisoned images, each embedded with a trigger that aligns with a specific target class, such as “Speed Limit”. As a result, each poisoned image can be misclassified as one of the targeted classes, depending on the specific trigger presented in the image.}
Our main contributions are summarized as follows:

  \begin{table*}[!t]
\centering
\setlength\tabcolsep{10pt}
\vspace{-3mm}
\caption{Properties of different backdoor attacks.}\label{Tab:MethodsComp}
\begin{threeparttable}
\begin{tabular}{l|l|cccc} 
\hline
\multicolumn{2}{c|}{Backdoor Attacks} & \makecell[c]{Input-Aware\\ \cite{nguyen2020input}}&  \makecell[c]{WaNet\\ \cite{nguyen2021wanet}} & \makecell[c]{One-to-N\\ \cite{xue2020one}} & Ours  \\  
\hline
\multirow{3}{*}{\makecell[c]{Characteristic\\ of Trigger}}& Invisible    & $\times$ & $\checkmark$  &     $\times$     &   $\checkmark$     \\
  & Dynamic & $\checkmark$ &$\checkmark$  & $\times$  &    $\checkmark$    \\
 & Clean Features\tnote{1} &$\times$ &$\times$ &$\times$&  $\checkmark$     \\
\hline 
\multirow{4}{*}{\makecell[c]{Resistance \\to defence}}  & Fine-Pruning~\cite{liu2018fine} & $\checkmark$ &$\checkmark$  &$\checkmark$&     $\checkmark$  \\
& Neural Cleanse~\cite{wang2019neural}  & $\checkmark$ & $\checkmark$  & $\checkmark$ &   $\checkmark$    \\
 & STRIP \cite{gao2019strip}   & $\checkmark$ &$\checkmark$  &\red{$\checkmark$}&    $\checkmark$   \\
& SentiNet~\cite{chou2020sentinet}  & $\checkmark$ & $\checkmark$ & $\times$ &      $\checkmark$ \\
\hline
\multirow{5}{*}{\makecell[l]{Robustness to \\Pre-processing\tnote{2}}}  & Flippling & Strong  & Strong & Strong & Strong  \\
                                                            &Rotation &Strong & Weak&Strong & Strong  \\
                                                            &Shrinking\&Padding  &Strong  & Strong & Weak & Strong  \\ 
                                                            &Cropping\&Resizing & Strong & Strong & Strong &  Strong \\
                                                             &Gaussian Noise Blurring & Strong & Weak & Strong  & Strong \\
\hline
\multicolumn{2}{l|}{Multiple Targets\tnote{3}}   &  $\times$   & $\times$  &    $\checkmark$       &   $\checkmark$   \\
\hline
\multicolumn{2}{l|}{Multiple Triggers\tnote{4} }    &  $\times$   & $\times$  &   $\times$     &   $\checkmark$ \\
\hline
\end{tabular}
    \begin{tablenotes}   
        \item[1] ``Clean Features'' means that the features of the triggers are from the training dataset.          
        \item[2] A backdoor attack is considered to have strong robustness to a pre-processing operation if the attack success rate (ASR) is larger than \red{50\% on the CIFAR-10 dataset}. 
        \item[3] \red{``Multiple Targets" enables attackers to manipulate a single input to generate various poisoned inputs targeting different target classes.}
        \item[4] \red{``Multiple Triggers'' in our attack denotes that each trigger can independently activate its corresponding backdoor. This differs from the N-to-One attack in~\cite{xue2020one}, where the backdoor is activated when all discrete triggers are simultaneously satisfied.}
      \end{tablenotes}   
\end{threeparttable}
 \vspace{-3mm}
\end{table*}




\begin{itemize}
\item We propose a new $M$-to-$N$ backdoor paradigm, which, to our best knowledge, is the first multi-target and multi-trigger attack. This paradigm enables an attacker to manipulate any clean image to generate different poisoned images for attacking multiple target classes simultaneously. The backdoor of each target class can be activated by any of multiple triggers, adding complexity to the detection and mitigation of the attack.

\item We develop a new strategy for generating poisoned images by
using clean images from each of the target classes as our triggers and embedding them in other clean images in an invisible manner. This strategy properly embeds trigger images in clean images, enabling attackers to launch an effective multi-target and multi-trigger attack by poisoning the training dataset with the trigger images from multiple target classes.

\item Extensive experimental results demonstrate that our $M$-to-$N$ attack is highly effective in simultaneously attacking multiple target classes when poisoning only a small portion of the training dataset (e.g. 2\%), and it exhibits high robustness against pre-processing operations and has high ability to resist existing defenses.
\end{itemize}

The rest of this paper is organized as follows. Section~\ref{sec: pre} reviews existing backdoor attacks and backdoor defenses.
Section~\ref{sec: threatandformulation} presents our threat model and problem formulation.
Section~\ref{Sec:Ourmethod} introduces our $M$-to-$N$ backdoor attack in detail. Section~\ref{sec: experiments} shows the experimental results, Section~\ref{sec: Resdefences} validates the ability of our attack to circumvent existing defenses, and Section~\ref{discussion} discusses the proposed attack. \red{Section~\ref{sec: conclude} concludes this paper.}

\section{Related Work}
\label{sec: pre}

\subsection{Backdoor Attacks}
\label{sec: backdoor}

Most backdoor attacks inject a backdoor into a DNN model by poisoning a small part of the training dataset with pre-defined triggers. Based on whether the labels of the poisoned images are changed or not, most poison-based backdoor attacks can be classified into two types: dirty-label backdoor attacks and clean-label backdoor attacks.

\subsubsection{Single-target Backdoor Attacks}
The majority of previous backdoor attacks focus on manipulating all the inputs with triggers to attack only a single target class~\cite{gong2023kaleidoscope,xu2023batt}, or manipulating inputs from a source class $i$ with triggers to attack the target class $i+1$ (inputs belonging to one source class can only be manipulated to attack a single target class, typically the subsequent class)~\cite{gu2017badnets,nguyen2021wanet}. As a result, the contamination of a specific sample can only attack a designated target class. For instance, Gu~\etal~\cite{gu2017badnets} proposed the first backdoor attack, which poisons a small part of the training dataset using a pre-defined trigger (e.g., a white square).
Apart from the white square in BadNets~\cite{gu2017badnets}, triggers are designed to be more complex patterns, such as dynamic and invisible patterns~\cite{nguyen2020input,li2021invisible,Gong2022ATTEQNNAQ,gong2021defense,gao2023not,gao2023backdoor,gong2023b3}, physical RGB filters~\cite{gong2023kaleidoscope} and asymmetric triggers~\cite{qi2023revisiting}, in order to evade existing defenses. All the poisoned images in the above attacks share the same trigger and a single target class to achieve a high ASR.

\subsubsection{Other Backdoor Attacks}
Different from these single-target attacks, more complex paradigms like untargeted and multi-target backdoor attacks offer varied functionalities.
Li~\etal~\cite{li2022untargeted} and Xue~\etal~\cite{xue2023untargeted} explored the untargeted backdoor attack, which involves altering the labels of poisoned images to random labels, resulting in poisoned models misclassifying poisoned images into arbitrary categories. While this paradigm does not allow for precise classification into an attacker-specified target class, it finds utility in applications like copyright protection of datasets~\cite{guo2023domain,li2023black,li2022move}.

In the realm of multi-target attacks, the One-to-N attack~\cite{xue2020one} and Marksman~\cite{doan2022marksman} methods are noteworthy. The One-to-N uses different levels of pixel modification inside a small square at the bottom right corner of the clean images as the triggers. The triggers are visible and can be easily detected by manual detection. Due to the simplicity of the trigger settings, the One-to-N attack exhibits lower ASRs on various datasets, particularly when the poisoned inputs undergo some common pre-processing operations. The Marksman method typically necessitates meticulous control over the training process, involving the simultaneous training of the original task model and optimizing trigger perturbations. Consequently, it exhibits certain limitations in terms of transferability across different model architectures. Besides, the Marksman method employs a single trigger for a target. In contrast, our attack does not require control over the training process, enabling the simultaneous use of multiple triggers to attack the same target category. Moreover, it achieves close to a 100\% attack success rate across different model architectures.

\subsection{Backdoor Defenses}
\label{sec: defence}
Since backdoor attacks present a serious security risk to DNN models, many backdoor defenses have been developed to detect and mitigate backdoor attacks. They can be roughly classified as
model-based defences~\cite{wang2019neural,liu2018fine,xiang2020detection,xiang2022post,li2021neural} and input-based defenses~\cite{gao2019strip,chou2020sentinet,LI20211,tran2018spectral,hayase2020spectre}. 

\subsubsection{Model-Based Defences}
Model-based defenses aim to detect and mitigate backdoors in a suspicious model. 
Liu~\etal~\cite{liu2018fine} and Wu~\etal~\cite{wu2021adversarial} proposed to use fine-pruning-based strategies to remove the trigger-related neurons from the suspicious models. 
In~\cite{liu2018fine}, malicious neurons are identified as dormant neurons with clean inputs, while in~\cite{wu2021adversarial}, malicious neurons are identified as neurons sensitive to adversarial perturbations. 
Knowledge distillation~\cite{li2021neural,gong2023redeem} and meta-learning techniques~\cite{xu2021detecting} were also used to detect backdoors.
The above backdoor defenses detect and mitigate possible attacks without knowing the triggers.
In contrast, the authors in~\cite{wang2019neural,xiang2022post} proposed to reverse-engineer the trigger of the target-class label and then eliminate the hidden backdoor by unlearning the relationship between the trigger and the target-class label.

\subsubsection{Input-Based Defenses}
The input-based defenses aim to prevent backdoor attacks by recognizing the poisoned inputs during either the training or testing process. 
Tran~\etal~\cite{tran2018spectral,hayase2020spectre} proposed to use the spectrum of the covariance to distinguish the poisoned images from the clean images. SentiNet~\cite{chou2020sentinet} uses the popular tool ``saliency map'' to locate the potential regions that may contain triggers. These defenses relied on full access to the model training process. Additionally, differences in feature similarity~\cite{jebreel2023defending} or learning rates between clean and poisoned samples are also used to filter malicious samples in the training set.

Some other input-based defenses aim to detect the poisoned inputs during the testing process.  
Gao~\etal~\cite{gao2019strip} proposed an input-based defense called STRIP that can filter the poisoned inputs from the clean inputs. It was proposed based on the insight that the predictions of perturbed clean inputs are random, whereas the predictions of poisoned inputs are fixed. Li~\etal~\cite{guo2023scale,xie2023badexpert} filtered malicious testing samples by analyzing their prediction consistency during the modification of the inputs and models process. Only the inputs certified as clean 
inputs can be fed into the models, preventing the poisoned inputs from activating the backdoors.

\section{Threat Model and Problem Formulation}
\label{sec: threatandformulation}

\subsection{Threat Model}
\subsubsection{Attacker's Goals}
\label{sec: goals}
Our attack considers a realistic threat scenario for the image classification task. The attacker expects to attain the following goals:

\noindent\textbf{Effectiveness.}
\red{The backdoor attack should achieve a high ASR, consistently predicting poisoned inputs as the target-class label(s) without degrading the performance on clean inputs, and should generalize across different DNN models.}


\begin{table}
\setlength{\tabcolsep}{1.6pt}
    \centering
     \caption{The important notations used in this paper.}
    \label{tab:notations}
    \begin{tabular}{ll}
    \toprule
        Notations & Descriptions \\
    \midrule
    $\bm{D}$ & The clean dataset and $\bm{D}= \bm{D}_{\textnormal{train}}\cup \bm{D}_{\textnormal{test}}$.\\
     $\bm{L}$ & The collection of attacker-specified target-class labels. \\ 
        $\bm{T}$ & The collection of triggers. \\
        $\bm{D}_b$ & The collection of poisoned samples. \\
        $\bm{D}_s$ & The collection of remaining clean samples in $\bm{D}_{\textnormal{train}}$.\\
        $\tilde{\bm{D}}_{\textnormal{train}}$ & The poisoned training dataset, $\tilde{\bm{D}}_{\textnormal{train}}=\bm{D}_b \cup \bm{D}_s$.\\
        $\bm{D}_L$ & The clean samples of the target classes in the $\bm{D}_{\textnormal{train}}$.\\
        \red{$\mathcal{H}$} & \red{\makecell[l]{The trigger embedding network that embeds a trigger into \\a clean image.}}\\
        \red{$\mathcal{R}$} & \red{\makecell[l]{The recovery network that extracts the trigger features from \\a poisoned image.}}\\
        \red{$\mathcal{D}$}& \red{\makecell[l]{The discriminator that identifies whether an input is a clean \\image or apoisoned image.}}\\
     \midrule
       $(x_i,y_i)$  & The $i$-th clean sample.\\
       $f_{\theta}(\cdot)$ & The clean model with parameters $\theta$.\\
        $f_{\tilde{\theta}}(\cdot)$ & The backdoored model with parameters $\hat{\theta}$.\\
        $l_k$ & The label of the $k$-th target class in $\bm{L}$.\\
        $t_j^{(k)}$ & The $j$-th trigger of the target class $l_k$. \\
        \red{$\hat{t}_j^{(k)}$} &  \red{The grayscale version of trigger $t_j^{(k)}$.}\\
        $(\tilde{x}_i,~\tilde{y}_i)$ & The poisoned sample generated from $(x_i,y_i)$.\\
        $G(x_i,t_j^{(k)})$ & The poisoned image generation function.\\
        $\rho$ & The poisoning ratio, i.e., $\rho= \frac{|\bm{D}_b|}{|\tilde{\bm{D}}_{\textnormal{train}}|}$.\\
    \bottomrule
    \end{tabular}
     \vspace{-3mm}
\end{table}


\noindent\textbf{Invisibility.}
\red{The triggers should be injected into clean images, making the poisoned images visually indistinguishable from their clean counterparts, thereby eluding manual detection.}

\noindent\textbf{Multiple targets.}
\red{The attack can simultaneously target multiple classes, causing the backdoored model to misclassify inputs into any of these specified target classes.}

\noindent\textbf{Multiple triggers.}
The backdoor of each target class can be formulated by multiple triggers, and any one of the triggers can activate the backdoor. 
A defender can only mitigate the attack by destroying all of the triggers.

\subsubsection{Attacker's Capabilities}
\red{We assume that attackers do not know the models' specific configurations, such as network structure, loss function, and optimizer. However, attackers have full access to the training datasets, a common assumption in backdoor attackers~\cite{gu2017badnets,xue2020one,li2021invisible,li2022untargeted}. }

\subsection{Problem Formalization}
We aim to poison a part of the training dataset by fusing triggers into some clean images so that the backdoors can be injected into the DNN models trained on the poisoned training dataset. Table~\ref{tab:notations} lists the important notations used in this paper.
Suppose that a dataset $\bm{D}=\{(x_i, y_i)|i=1,\cdots,N^{(D)}\}$ has $N^{(D)}$ i.i.d. clean samples and is comprised of the training dataset $\bm{D}_{\textnormal{train}}$ and the testing dataset $\bm{D}_{\textnormal{test}}$. 
The $i$-th clean sample comprises the clean image $x_i \in \bm{X}=\{0,...,255\}^{c\times w \times h}$ and its ground truth label $y_i\in \bm{Y}=\{1,...,C\}$. 
A poisoned image $\tilde{x}_i$ can be generated from the clean image $x_i$ using a poisoned image generation function. 
Thus, a poisoned sample $(\tilde{x}_i,\tilde{y}_i)$ can be generated by mapping the poisoned image $\tilde{x}_i$ to the target-class label $\tilde{y}_i$. Attackers can obtain a poisoned training dataset $\tilde{\bm{D}}_{\textnormal{train}}$ by using the poisoned samples to replace the corresponding clean samples in $\bm{D}_\textnormal{train}$, namely $\tilde{\bm{D}}_{\textnormal{train}}=\bm{D}_b\cup \bm{D}_s$, where $\bm{D}_b$ is the set of poisoned samples and $\bm{D}_s$ is the set of remaining clean samples in $\bm{D}_\textnormal{train}$. The $\rho=|\bm{D}_b|/|\tilde{\bm{D}}_{\textnormal{train}}|$ is the poisoning ratio within the poisoned training dataset. 

Prior backdoor attacks can threaten only one target class, and the backdoor can be activated by a fixed trigger. Our $M$-to-$N$ backdoor attack can attack $N$ target classes simultaneously, and the backdoor of each target class can be activated by any one of $M$ triggers. 
We assume that the labels of $N$ target class to be attacked are $\bm{L}=\{l_k|l_k\in \bm{Y},~k=1,\cdots,\textit{N}\}$. Since each target class has $M$ triggers, we present the $M\times N$ triggers as $\bm{T}=\{t_j^{(k)}|k=1,\cdots,\textit{N};j=1,\cdots,\textit{M}\}$, where $t_1^{(k)},t_2^{(k)},\cdots,t_M^{(k)}$ are the $M$ triggers of the $k$-th target class, and they are randomly selected from the target-class images in $\bm{D}_{\textnormal{train}}$. 

A poisoned sample for the $k$-th target class is generated as follows. (1) Select a trigger $t_j^{(k)}$ from the $M$ triggers of the $k$-th target class and generate a poisoned image $\tilde{x}_i$ by embedding $t_j^{(k)}$ in a clean image $x_i$ that does not correspond to the $N$ target classes, namely 
$x_i \in \bm{D}_{\textnormal{train}}\verb|\|\bm{D}_{L}$. 
The poisoned image generation can be presented as $\tilde{x}_i=G(x_i,t_j^{(k)})$. (2) Mark the poisoned image $\tilde{x}_i$ as the target-class label $\tilde{y}_i$ and obtain a poisoned sample $(\tilde{x}_i,\tilde{y}_i)$ for the $k$-th target class $l_k$.
Since there are $M$ triggers $t_1^{(k)},t_2^{(k)},\cdots,t_M^{(k)}$ corresponding to the target-class label $l_k$, we can generate the poisoned samples to attack the $k$-th target class by uniformly using these $M$ triggers. Similarly, the poisoned samples for all the $N$ attacked target classes can be generated. All the poisoned samples form the poisoned samples set $\bm{D}_{b}$. 

The behavior of the backdoored model $f_{\hat{\theta}}(\cdot)$ can be formulated as: 
\begin{equation}
    f_{\hat{\theta}}(x)=f_{\theta}(x)=y \wedge f_{\theta}(\tilde{x})=y \wedge f_{\hat{\theta}}(\tilde{x})=\tilde{y},
\end{equation}
where $f_{\theta}(\cdot)$ is the clean model, $(x,y)\in \bm{D}_s$ is a benign sample and $(\tilde{x},\tilde{y})\in \bm{D}_b$ is a poisoned sample. For each clean image $x$, both the clean model $f_{\theta}(\cdot)$ and backdoored model $f_{\hat{\theta}}(\cdot)$ can output the true label $y$. For the poisoned image $\tilde{x}_i$, the clean model $f_{\theta}(\cdot)$ outputs the real label $y$; however, the backdoored model $f_{\hat{\theta}}$ outputs the target-class label $\tilde{y}_i$, which exhibits the attacker's expected behaviors.

\section{Our Method}
\label{Sec:Ourmethod}

\red{In this section, we introduce our $M$-to-$N$ backdoor attack in detail. We assume that an attacker would like to inject backdoors for $N$ target classes.
For each target class, the attacker first selects $M$ target-class clean images as the $M$ triggers of the target class and then develops a poisoned image generation framework by using all the $M\times N$ triggers to poison a part of the training dataset equally.  
A DNN model will be injected backdoors when it is trained on the poisoned training dataset. The backdoored model can output any attacker's expected target-class label when the input is poisoned by a corresponding trigger. }




\begin{figure*}
\centering
    \begin{minipage}[b]{0.95\linewidth}    
    \centering
        \includegraphics[width=0.99\textwidth]{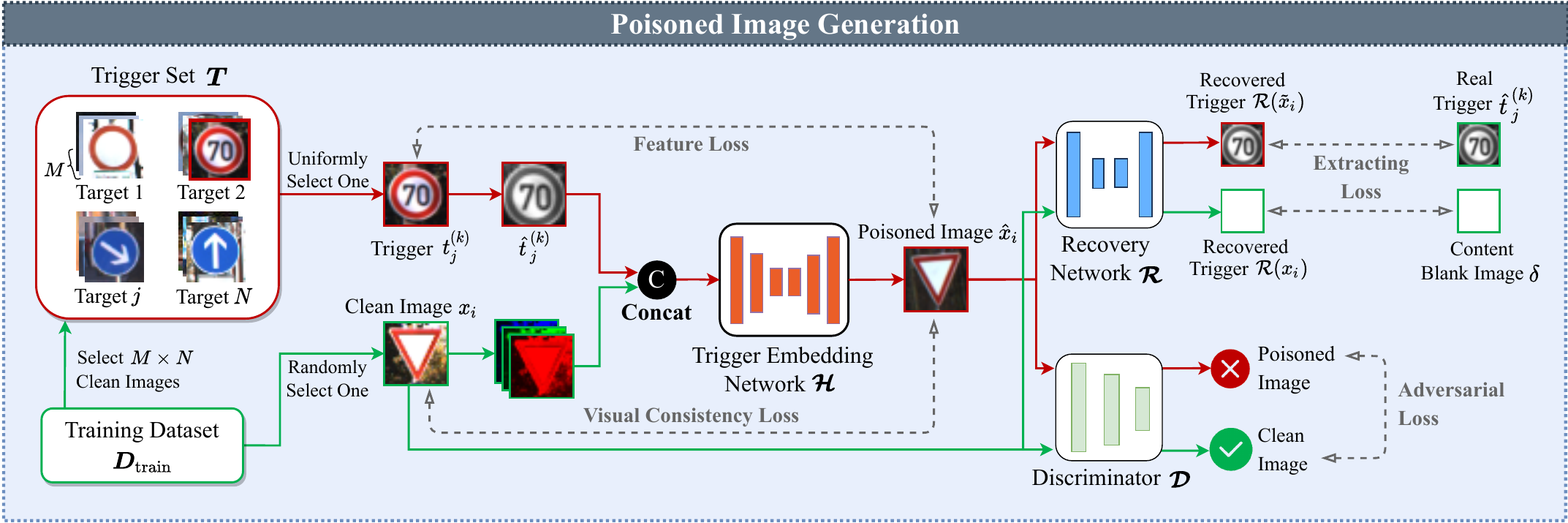}
    \caption{The schematic diagram of our poisoned image generation framework.}
    \label{fig:poisonedgeneration}
    \end{minipage}
     \vspace{-3mm}
\end{figure*}

\subsection{Trigger Selection}
\red{We randomly select $M$ clean images corresponding to each target class in the training dataset and use their grayscale versions as the $M$ triggers, which can achieve the following advantages.}
\begin{itemize}
\item \red{The triggers have the same distribution as the clean images in the training dataset, enabling the DNN models to generalize to these triggers as the training progresses, thereby ensuring the effectiveness of our attack across multiple target classes.}

\item The triggers are comprised of clean features and cannot bring extra features that are nonexistent in the training dataset. Thus, the backdoored models are difficult to be detected by existing defenses such as~\cite{wang2019neural}, according to the discussions in~\cite{lin2020composite}. 

\end{itemize}

Converting trigger images into grayscale format can simplify the input space and reduce the computational complexity during poisoned image generation, as an RGB color image has three channels while a grayscale image has only one.
Additionally, the conversion operation can prevent DNN models from being influenced by color information, making the attack more effective in multi-target and multi-trigger scenarios.

\subsection{Poisoned Image Generation Framework}
We design a poisoned image generation framework to embed our triggers in other clean images. Its structure is shown in Fig.~\ref{fig:poisonedgeneration}, which is developed according to the DNN-based watermarking in~\cite{zhang2021deep,9222304}. As can be seen, the generation framework comprises three networks: trigger embedding network $\mathcal{H}$, recovery network $\mathcal{R}$, and discriminator $\mathcal{D}$.
$\mathcal{H}$ is an encoder-decoder network that embeds a trigger into a clean image, and the encoded result is directly used as our poisoned image. To ensure embedding quality, the poisoned image is then separately sent into $\mathcal{R}$ and $\mathcal{D}$. We assume that a trigger is successfully injected into a poisoned image if it can be extracted from the poisoned image. Thus, we use $\mathcal{R}$ to extract the trigger features from the poisoned image. $\mathcal{D}$ is used to identify whether an input is a clean image or a poisoned image, and we use it to assist $\mathcal{H}$ in generating high-quality poisoned images that are indistinguishable from clean images.

We train the generation framework on the training dataset $\bm{D}_{\textnormal{train}}$. Specifically, we first randomly select a trigger $t_j^{(k)}$ from the trigger set $\bm{T}$ and a clean image $x_i$ from the training dataset $\bm{D}_{\textnormal{train}}$ as the inputs of the generation framework.
We then concatenate the clean image $x_i$ with the grayscale version $\hat{t}_j^{(k)}$ of the trigger along the channel dimension and feed the composite image into the trigger embedding network $\mathcal{H}$.
The output of $\mathcal{H}$ is a poisoned image $\tilde{x}_i$ and assigned with the target-class label $l_k$, i.e., 
\begin{equation}
\label{eq: concat}
    \tilde{x}_i=\mathcal{H}([x_i;\hat{t}_j^{(k)}]), \quad  \tilde{y}_i=l_k,
\end{equation}
where $x_i\in \bm{D}_{\textnormal{train}}$, and $l_k\in \bm{L}$. 




\subsubsection{Network Structures}
We use UNet~\cite{ronneberger2015u} as the network structure of $\mathcal{H}$ because it performs well in situations where the inputs and outputs share some common properties. We modify the stride of each convolution layer from $2$ to 1 to apply to images of small size (e.g., $32\times 32$). 
Besides, we use the widely-used CEILNet~\cite{fan2017generic} and PatchGAN~\cite{isola2017image} as the network structures of $\mathcal{R}$ and $\mathcal{D}$, respectively.

\subsubsection{Loss Functions}
The overall loss function of the poisoned image generation framework comprises the losses $L_H$, $L_{R}$ and $L_{D}$  of the above three networks, namely 
\begin{equation}
\label{eq: sumloss}
    L = \lambda_{H} L_{H} + \lambda_R L_R  +  \lambda_{D} L_{D},
\end{equation}
where $\lambda_{H}$, $\lambda_{R}$, and $\lambda_{D}$ are the hyper-parameters to balance the three loss terms.

\noindent\textbf{Embedding Loss $L_{H}$.}
The embedding Loss $L_{H}$ comprises the visual consistency part $L_{V}$ and feature part $L_{F}$, i.e.,
\begin{equation}
\label{eq: hidden}
    L_H = \lambda_{H}^{(1)} L_{V} + \lambda_{H}^{(2)} L_F,
\end{equation}
\red{where $\lambda_{H}^{(1)}$ and $\lambda_{H}^{(2)}$ are the hyper-parameters.}

The visual consistency loss evaluates the differences between poisoned and clean images. It is used to help the trigger embedding network $\mathcal{H}$ to generate poisoned images that are indistinguishable from the related clean images. We evaluate $L_V$ by calculating the pixel distance between the poisoned image $\tilde{x}_i$ and the clean image $x_i$, namely
\begin{equation}
\label{l2loss}
    L_{V} = \mathop{\mathbb{E}}_{x_i\in \bm{D}_{\textnormal{train}}}[||\tilde{x}_i-x_i||^2]. 
\end{equation}
Similar to prior DNN-based encoding works in~\cite{zhang2021deep,9663305}, we also calculate the pixel distance using $L_2$ Norm. 

The feature loss evaluates the feature distance between the poisoned images and the triggers, ensuring that the poisoned images contain the semantics of triggers.
We use a DNN-based feature extractor introduced in~\cite{johnson2016perceptual} to extract the features of the poisoned images and triggers, and $ L_{F}$ is the difference between the two features, namely 
\begin{equation}
\label{vggloss}
    L_{F} = \mathop{\mathbb{E}}_{x_i\in \bm{D}_{\textnormal{train}}}\frac{1}{N_k}[||V_k(\tilde{x}_i)-V_k(t_j^{(k)})||^2],
\end{equation}
where $V_k(\cdot)$ is the features extracted by the $k$-th layer of VGG model and $N_k$ denotes the number of feature neurons.

\noindent\textbf{Extracting Loss $L_R$.}
We use $L_R$ to guide the recovery network $\mathcal{R}$ to extract the trigger information from the poisoned images.
$L_R$ evaluates the pixel difference between the recovered hidden information $R(\tilde{x}_i)$ and the real information $\hat{t}_j^{(k)}$.
To prevent the recovery network $\mathcal{R}$ from overfitting the poisoned images, we also input the clean image $x_i$ into $\mathcal{R}$ to evaluate the pixel difference between the $R(x_i)$ and the blank image $\delta$. Then the extracting loss $L_R$ is formulated as 

\begin{equation}
\label{lr}
    L_{R} = \mathop{\mathbb{E}}_{\begin{subarray}{l} \\
x_i\in \bm{D}_{\textnormal{train}}
\end{subarray}}[||\mathcal{R}(\tilde{x}_i)-\hat{t}_j^{(k)}||^2] + \mathop{\mathbb{E}}_{x_i\in \bm{D}_{\textnormal{train}}}[||\mathcal{R}(x_i)-\delta||^2].
\end{equation}

\noindent\textbf{Adversarial Loss $L_D$.}
We use the discriminator $\mathcal{D}$ to determine whether an input is a clean or poisoned image. $L_{D}$ can assist $\mathcal{H}$ in embedding triggers properly such that $\mathcal{D}$ can not distinguish the poisoned images from real clean images in $\bm{D}_{\textnormal{train}}$.
It is formulated as
\begin{equation}
\label{ganloss}
    L_{D} = \mathop{\mathbb{E}}_{\begin{subarray}{l}x_i\in \bm{D}_{\textnormal{train}}
\end{subarray}}[\log(1-\mathcal{D}(\tilde{x}_i))] + \mathop{\mathbb{E}}_{x_i\in \bm{D}_{\textnormal{train}}}[\log(\mathcal{D}(x_i))].
\end{equation}

Finally, Algorithm~\ref{algHDR} presents the pseudo-code of the training details of the three networks $\mathcal{H}$, $\mathcal{D}$, and $\mathcal{R}$ during the poisoned image generation process.

\begin{figure}[!t]
    \centering
\begin{minipage}[b]{0.99\linewidth}    
    \includegraphics[width=1\textwidth]{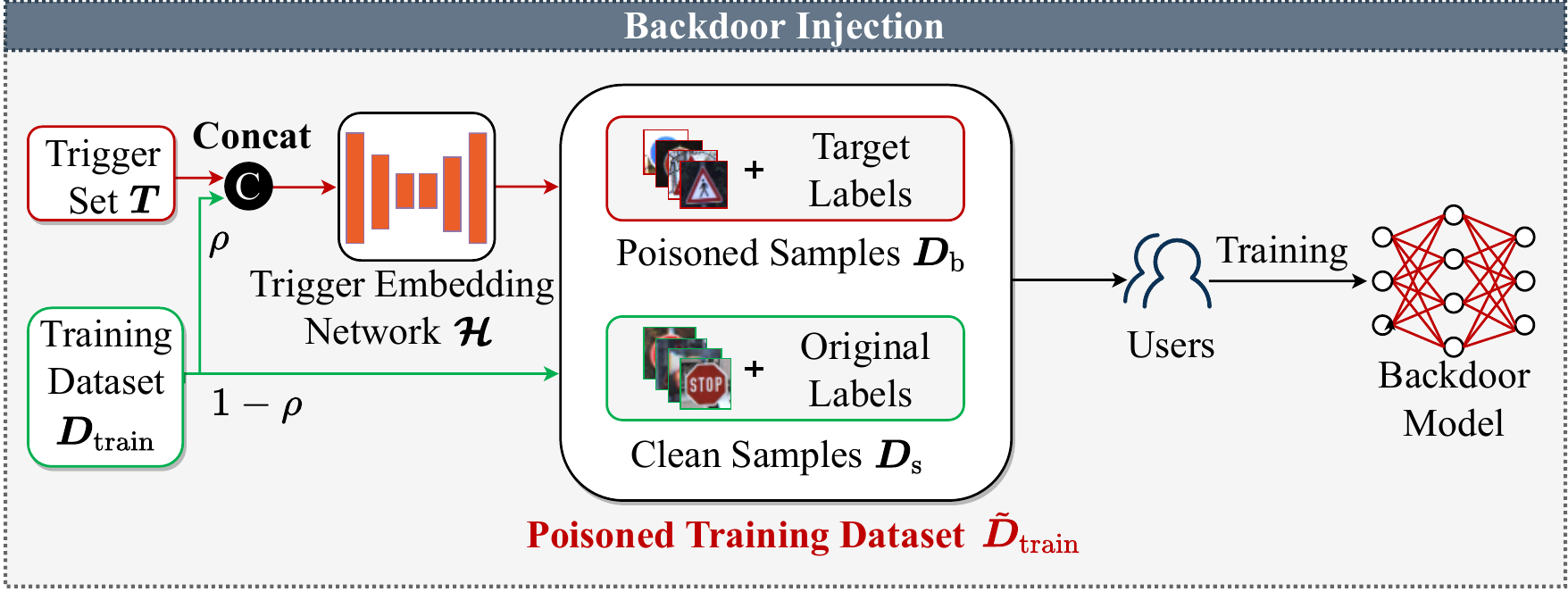}
    \caption{The process of our backdoor injection.}
    \label{fig:backdoorinjection}
    \end{minipage}
\end{figure}

\begin{algorithm}[!t]
\caption{The training process of \(\mathcal{H}\), \(\mathcal{R}\) and \(\mathcal{D}\).}
\label{algHDR}
\begin{algorithmic}
\REQUIRE
\begin{enumerate}
    \item The initial parameters: \(\theta_H\), \(\theta_R\) and \(\theta_D\); 
    \item The training dataset: \(\bm{D}_{\textnormal{train}}\);
    \item The triggers set: \(\bm{T}\);
    \item The loss weights: \(\lambda_H\), \(\lambda_R\), \(\lambda_D\), \(\lambda_{H}^{(1)}\), \(\lambda_{H}^{(2)}\);
    \item The learning rates: \(lr_H\), \(lr_R\), \(lr_D\);
    \item The number of iterations: \(I\);
    \item The blank image: \(\delta\).
\end{enumerate}
\ENSURE The optimized parameters: \(\theta_H\), \(\theta_R\) and \(\theta_D\);

\FOR{\(i \leftarrow 1\) \TO \(I\)}
    \FOR{each minibatch \((x, y) \subset \bm{D}_{\textnormal{train}}\)}
        \STATE Sample a minibatch of triggers \(t\) from \(\bm{T}\);
        \STATE Generate poisoned images \(\tilde{x}\) using Eq.~\eqref{eq: concat};
        \STATE Compute \(L_V (x, \tilde{x})\) using Eq.~\eqref{l2loss};
        \STATE Compute \(L_{F}(\tilde{x}, t)\) using Eq.~\eqref{vggloss};
        \STATE Compute \(L_{R}(\tilde{x}, \hat{t}, x, \delta)\) using Eq.~\eqref{lr};
        \STATE Compute \(L_{D}(x, \tilde{x})\) using Eq.~\eqref{ganloss};
        \STATE \(L_H = \lambda_{H}^{(1)} L_{V} + \lambda_{H}^{(2)} L_{F}\); 
        \STATE \(L = \lambda_{H} L_{H} + \lambda_R L_R +  \lambda_{D} L_{D}\);
        \STATE \(\theta_H \leftarrow \theta_H - lr_H \nabla_{\theta_H} L\);
        \STATE \(\theta_R \leftarrow \theta_R - lr_R \nabla_{\theta_R} L_R\);
        \STATE \(\theta_D \leftarrow \theta_D - lr_D \nabla_{\theta_D} L_D\);
    \ENDFOR
\ENDFOR
\end{algorithmic}
\end{algorithm}

\subsection{Backdoor Injection}
Fig.~\ref{fig:backdoorinjection} shows the backdoor injection process, which includes generating a poisoned training dataset and training the DNN model on the dataset. The attacker poisons a part of the training dataset $\bm{D}_{\textnormal{train}}$ to obtain the poisoned training dataset $\tilde{\bm{D}}_{\textnormal{train}}$, and the backdoors can be secretly implanted into a DNN model trained normally on  $\tilde{\bm{D}}_{\textnormal{train}}$. 


When generating the poisoned training dataset, the attacker uses the well-trained trigger embedding network $\mathcal{H}$ to poison a part of the training dataset with the poisoning ratio $\rho$. Specifically, the attacker randomly selects $\rho$ clean images from the dataset $\bm{D}_{\textnormal{train}}$\verb|\|$\bm{D}_{L}$, and poisons each of the selected clean image using the following steps. (1) Select a trigger uniformly from the trigger set $\bm{T}$ and convert it into a grayscale image. (2) Concatenate the clean image with the grayscaled trigger. (3) Use the concatenated result as the input of $\mathcal{H}$ to generate a poisoned image. (4) A poisoned sample is obtained by marking the poisoned image as the target-class label that corresponds to the used trigger.
After poisoning all the selected clean images, the attacker can get a poisoned training dataset $\tilde{\bm{D}}_{\textnormal{train}}$ by replacing the original clean samples in $\bm{D}_\textnormal{train}$ with the generated poisoned samples.
Since a uniformly selected trigger is used to poison a clean image, the generated poisoned images contain the semantics of all the $M \times N$ triggers for the $N$ attacker-specified target classes. 

Once the poisoned dataset $\tilde{\bm{D}}_{\textnormal{train}}$ is obtained, the backdoors can be implanted into a DNN model when a user trains the model on $\tilde{\bm{D}}_{\textnormal{train}}$, and finally a backdoored model $f_{\hat{\theta}}$ is obtained as
\begin{equation} \label{eqtrain}
\min\limits_{\theta}\mathop{\mathbb{E}}_{(x,y)\in \tilde{\bm{D}}_{\textnormal{train}}}L_c(f_{\hat{\theta}}(x),y),
\end{equation}
where $L_c$ is the loss function of the DNN model. 




\section{Experimental Results}
\label{sec: experiments}

\subsection{Implementation Settings}
\label{setup}
We conduct backdoor attack experiments on MNIST using MNIST\_CNN model~\cite{nguyen2021wanet} and Mobilenetv2~\cite{sandler2018mobilenetv2}, CIFAR-10 and GTSRB datasets with PreActRes18~\cite{he2016identity}, Resnet18~\cite{he2016deep} models, and a subset of ImageNet (ImageNet-10) using a pre-trained Resnet18 model with a fully connected layer. The top 10 classes with the highest image count from ImageNet are selected.

\red{To evaluate the effectiveness of the attacks, we utilize three metrics: Attack Success Rate (ASR), Backdoored Model Accuracy (BA), and Clean Model Accuracy (CA), consistent with the settings in~\cite{nguyen2021wanet,nguyen2020input,xue2020one}. ASR assesses how effectively poisoned inputs are classified into target classes; BA measures the accuracy of a backdoored model on a clean testing dataset; and CA measures the accuracy of a clean model on a clean testing dataset, where a smaller difference between BA and CA indicates minimal impact on the model’s original task.}

\red{Additionally, we measure the visual quality of poisoned images using peak-signal-to-noise-ratio (PSNR), structural similarity index measure  (SSIM), and learned perceptual image patch similarity (LPIPS), where higher PSNR and SSIM, and lower LPIPS suggest more natural image appearances~\cite{9903662}.}

The trigger embedding network $\mathcal{H}$ is trained with the Adam optimizer, which starts with a learning rate of 0.0002, and its learning rate will decay by 0.2 if the loss does not decrease within five epochs. DNN models are trained with the SGD optimizer, whose initial learning rate is 0.01, and reduced by a factor of 10 when the training epochs reach 100 and 150. The maximum epoch is 200. The hyper-parameters $\lambda_{H}$, $\lambda_{H}^{(1)}$, $\lambda_{H}^{(2)}$, and $\lambda_R$ are 1, and $\lambda_{D}$ is 0.01. All experiments are performed on a server with the Ubuntu 16.04.6 LTS operating system, equipped with a 3.20GHz CPU, NVIDIA's GeForce GTX3090 GPU with 62G RAM, and an 8TB hard disk.

\begin{table*}[!t]
\small
\caption{Experimental results (\%) of the $M$-to-$N$ attack on MNIST, CIFAR-10, and GTSRB datasets with different models.}
\label{table:oureffectiveness}
\centering
\setlength{\tabcolsep}{3pt}
\begin{tabular}{ccccccccccccccccccc}        
\toprule
Datasets (Models)& \multicolumn{5}{c}{MNIST (MNIST\_CNN)~\cite{nguyen2021wanet}} & \multicolumn{5}{c}{CIFAR-10 (PreActRes18~\cite{he2016identity})} & \multicolumn{5}{c}{GTSRB (PreActRes18~\cite{he2016identity})} \\
\cmidrule(r){1-1}\cmidrule(lr){2-6} \cmidrule(lr){7-11} \cmidrule(l){12-16}
$M \rightarrow$ & &\multicolumn{2}{c}{1} & \multicolumn{2}{c}{10} & &\multicolumn{2}{c}{1} & \multicolumn{2}{c}{10} & &\multicolumn{2}{c}{1} & \multicolumn{2}{c}{10}\\ \cmidrule(lr){3-4}\cmidrule(lr){5-6} \cmidrule(lr){8-9}\cmidrule(lr){10-11} \cmidrule(lr){13-14}\cmidrule(l){15-16}
$N \downarrow$ & CA & BA & ASR & BA & ASR   & CA& BA & ASR & BA& ASR  & CA& BA  & ASR& BA& ASR  \\
\cmidrule(lr){2-2}\cmidrule(lr){3-4}\cmidrule(lr){5-6}  \cmidrule(lr){7-7}  \cmidrule(lr){8-9} \cmidrule(lr){10-11}  \cmidrule(lr){12-12}  \cmidrule(lr){13-14} \cmidrule(l){15-16}
1 & \multirow{3}{*}{99.52} & 99.36 & 100 & 99.4 & 100 &\multirow{4}{*}{94.69} &94.42 & 99.76  &93.97 & 99.97& \multirow{4}{*}{99.40} &99.38 & 100  & 99.20 & 99.91 \\
2 & & 98.93 & 99.50 & 99.29 & 100&   & 94.43 &  98.51 & 93.83 & 99.43 & & 99.36 & 100  & 99.08 & 98.88\\
4 &  & 99.36 & 99.93 & 99.43 & 96.24& &94.38 & 99.70  & 93.49 & 98.23 & & 99.36 & 99.89 & 99.06 & 99.13\\
5 & &99.44 & 95.06 & 99.40 & 95.49 &  & 94.44&  99.79 &93.94 & 99.88 & &99.25 & 99.61 & 99.44 & 99.77\\
\hline
\hline
Datasets (Models)& \multicolumn{5}{c}{MNIST (Mobilenetv2~\cite{sandler2018mobilenetv2})} & \multicolumn{5}{c}{CIFAR-10 (Resnet18~\cite{he2016deep})} & \multicolumn{5}{c}{GTSRB (Resnet18~\cite{he2016deep})} \\
\cmidrule(r){1-1}\cmidrule(lr){2-6} \cmidrule(lr){7-11} \cmidrule(l){12-16}
$M \rightarrow$ & &\multicolumn{2}{c}{1} & \multicolumn{2}{c}{10} & &\multicolumn{2}{c}{1} & \multicolumn{2}{c}{10} & &\multicolumn{2}{c}{1} & \multicolumn{2}{c}{10}\\
\cmidrule(lr){3-4}\cmidrule(lr){5-6} \cmidrule(lr){8-9}\cmidrule(lr){10-11} \cmidrule(lr){13-14}\cmidrule(l){15-16}
$N \downarrow$ & CA & BA & ASR & BA & ASR   & CA& BA & ASR & BA& ASR  & CA& BA  & ASR& BA& ASR  \\
\cmidrule(lr){2-2}\cmidrule(lr){3-4}\cmidrule(lr){5-6}  \cmidrule(lr){7-7}  \cmidrule(lr){8-9} \cmidrule(lr){10-11}  \cmidrule(lr){12-12}  \cmidrule(lr){13-14} \cmidrule(l){15-16}
 1 & \multirow{4}{*}{99.50} & 99.25 & 100 & 99.24 & 100 & \multirow{4}{*}{94.02} &93.23 & 99.97  & 93.68 & 99.97 &\multirow{4}{*}{99.42} & 99.40 & 100  & 99.22 & 99.90\\
                         2 & &99.26 & 100 & 99.24 & 99.74 & & 93.11 &  96.75 & 93.70 & 99.38 & & 99.47 & 100   &  99.33 & 98.24\\
                         4 & &99.27 & 100 & 99.18 & 99.97& & 93.02 & 93.62  & 93.35 & 95.55 & &99.04 & 99.78 & 99.21 & 98.47\\
                         5 & &99.1 & 99.81 &  99.17 & 98.51& & 93.32&  99.72 & 93.87 & 99.81 & &99.22 &  99.64  & 99.35 & 99.89 \\
\bottomrule
\end{tabular}
\vspace{-3mm}
\end{table*}

\begin{table}[!t]
\caption{Experimental results of our $M$-to-$N$ backdoor attack on the ImageNet-10 dataset with Resnet18.}
\label{table:oureffectivenessimagenet}
\centering
\begin{tabular}{ccccccc}         
\toprule
$M \rightarrow$ & \multicolumn{2}{c}{1}& \multicolumn{2}{c}{2}&\multirow{2}{*}{CA (\%)} \\
\cmidrule(l){2-3} \cmidrule(l){4-5}
 $N \downarrow$&BA (\%)& SR (\%) &BA (\%) & ASR (\%)\\ 
\midrule
1 & 99.20& 99.80& 98.40 & 99.80& \multirow{4}{*}{99.01}\\
 2 & 98.40 & 6.20 & 97.40 & 93.20 &\\
4 & 98.20& 94.20& 97.40 & 96.80&\\
5& 97.00 & 97.20 & 96.60 & 98.80 &\\
\bottomrule
\end{tabular}
\end{table}

\subsection{Attack Effectiveness}
\label{effectiveness}
\noindent\textbf{Our Attack Effectiveness.} We design experiments by setting $M\in\{1,10\}$, $N\in\{1,2,4,5\}$ on the MNIST, CIFAR-10, and GTSRB datasets.
To alleviate the pressure on the DNN model's generalization ability on high-resolution datasets (e.g., ImageNet), we set the number of triggers $M$ ($M \in \{1,2\}$). We employ the $M\times N$ triggers to uniformly poison the testing set, upon which we calculate the BA and ASR of the backdoored models. \red{Table~\ref{table:oureffectiveness} and Table~\ref{table:oureffectivenessimagenet} show that the ASRs for our $M$-to-$N$ attack approach 100\% across different values of $M$ and $N$, with BAs and CAs discrepancie around 1\%. The results validate the effectiveness of all triggers in activating the backdoors and the capability of our attack to target multiple targets with different triggers. Additionally, the uniformly high ASRs across various model architectures demonstrate the generalizability of our $M$-to-$N$ attack to different datasets and DNN models.}


\begin{table}[!t]
\centering
\caption{Comparison (\%) of our $M$-to-$N$ attack with advanced attacks against a single target class. \red{The highest BA and ASR scores are highlighted in bold.}}
\label{tab:comparedall}
\begin{tabular}{lcccccc} 
\toprule
\multirow{2}{*}{Datasets}  & \multicolumn{2}{c}{Input-Aware\cite{nguyen2020input}}  & \multicolumn{2}{c}{WaNet\cite{nguyen2021wanet}}  &  \multicolumn{2}{c}{Ours} \\ 
\cmidrule(lr){2-3}  \cmidrule(lr){4-5}  \cmidrule(l){6-7}
& BA   & ASR   & BA  & ASR & BA   & ASR\\
 \midrule
MNIST  & \textbf{99.54} & 99.54 & 99.52 &  99.86 &99.36 & \textbf{100}\\
CIFAR-10 & \textbf{94.65}  & 99.32  & 94.15           & \textbf{99.86}  & 94.42  & 99.76  \\
GTSRB &99.27& 99.84  &98.97 &  98.78 & \textbf{99.38}& \textbf{100}\\
ImageNet-10  & 97.80& 99.60 & 97.60 & 70.60  &  \textbf{99.20}&  \textbf{99.80} \\
\bottomrule
\end{tabular}
\vspace{-1.5em}
\end{table}    

\begin{figure*}[!t]
	\centering
	\begin{minipage}[b]{0.98\linewidth}
		\begin{minipage}[b]{0.105\linewidth}
            \centerline{\includegraphics[width=1\linewidth]{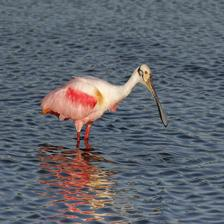}}
            \vspace{1mm}
               \centerline{\includegraphics[width=1\linewidth]{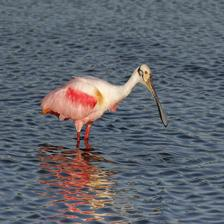}}
			\centerline{(a)}
		\end{minipage}
		\begin{minipage}[b]{0.105\linewidth}
                \centerline{\includegraphics[width=1\linewidth]{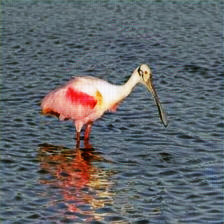}}
			\vspace{1mm}
               \centerline{\includegraphics[width=1\linewidth]{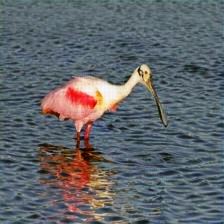}}
            \centerline{(b)}
		\end{minipage}
            \begin{minipage}[b]{0.105\linewidth}
    \centerline{\includegraphics[width=1\linewidth]{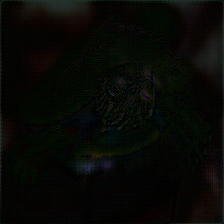}}
			\vspace{1mm}
               \centerline{\includegraphics[width=1\linewidth]{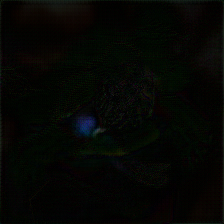}}
			\centerline{(c)}
		\end{minipage}
            \begin{minipage}[b]{0.105\linewidth}
            \centerline{\includegraphics[width=1\linewidth]{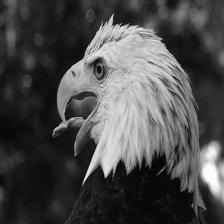}}
			\vspace{1mm}
               \centerline{\includegraphics[width=1\linewidth]{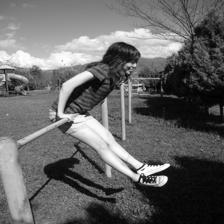}}
			\centerline{(d)}
		\end{minipage}
          \begin{minipage}[b]{0.105\linewidth}
           \centerline{\includegraphics[width=1\linewidth]{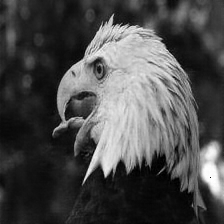}}
			\vspace{1mm}
               \centerline{\includegraphics[width=1\linewidth]{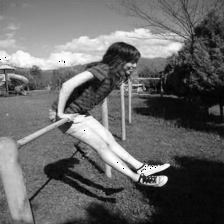}}
                \centerline{(e)}
		\end{minipage}
  		\hfill\vline\hfill
		\begin{minipage}[b]{0.105\linewidth}
            \centerline{\includegraphics[width=1\linewidth]{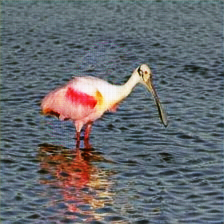}}
			\vspace{1mm}
               \centerline{\includegraphics[width=1\linewidth]{myimagenet_container_img_0_1.png}}
            \centerline{(f)}
		\end{minipage}
        \begin{minipage}[b]{0.105\linewidth}
            \centerline{\includegraphics[width=1\linewidth]{myimagenet_diff_0_1.png}}
			\vspace{1mm}
               \centerline{\includegraphics[width=1\linewidth]{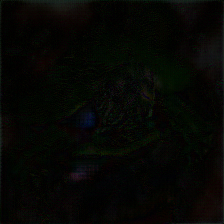}}
			\centerline{(g)}
		\end{minipage}
		\begin{minipage}[b]{0.105\linewidth}
             \centerline{\includegraphics[width=1\linewidth]{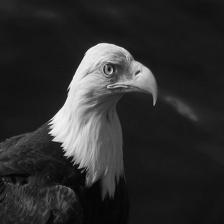}}
			\vspace{1mm}
               \centerline{\includegraphics[width=1\linewidth]{myimagenet_secret_img_0_1.png}}
			\centerline{(h)}
		\end{minipage}
		\begin{minipage}[b]{0.105\linewidth}
            \centerline{\includegraphics[width=1\linewidth]{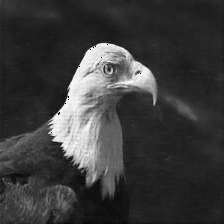}}
			\vspace{1mm}
               \centerline{\includegraphics[width=1\linewidth]{myimagenet_rev_secret_img_0_1.png}}
			\centerline{(i)}
		\end{minipage}
	\end{minipage}
	\caption{Visual effects of the poisoned images in our $M$-to-$N$ backdoor attack (1-to-2 attack on the left and 2-to-1 attack on the right). (a) Original images; (b) poisoned images for two target classes; (c) residuals of (a) and (b); (d) real triggers for generating (b); (e) recovered triggers from (b); (f) poisoned images for one target class using two different triggers; (g) residuals of (a) and (f); (h) real triggers for generating (f); (i) recovered triggers from (f).}
 \label{fig:visual}
 \vspace{-4mm}
\end{figure*}

\noindent\textbf{Comparison with Advanced Attacks.} To assess the effectiveness of the $M$-to-$N$ backdoor attack on a single target class, we conduct a comparison with two state-of-the-art backdoor attacks, 
Input-Aware~\cite{nguyen2020input} and WaNet~\cite{nguyen2021wanet}. \red{The comparison experiments are carried out using their publicly available source codes with the models described in the upper half of Table~\ref{table:oureffectiveness}.
Table~\ref{tab:comparedall} demonstrates that our $M$-to-$N$ attack achieves the highest ASRs on the MNIST, GTSRB, and ImageNet-10 datasets and exhibits modest performance on CIFAR-10. The negligible differences in BAs compared to Input-Aware are likely due to the inherent randomness in DNN training processes. Moreover, our attack maintains more consistent performance across different datasets than WaNet, which experiences a notable decline in ASR on ImageNet-10. }

\noindent\textbf{Comparison with the One-to-N attack.} \red{To evaluate the effectiveness of the $M$-to-$N$ attack on multiple target classes, we compare it with the One-to-N attack~\cite{xue2020one} across four datasets, targeting four classes simultaneously. Due to the unavailability of the source code for the One-to-N attack, our implementation may have variations in hyper-parameters and configuration settings. Generally, our attack outperforms the One-to-N attack in both BAs and ASRs, except for a slight 0.4\% lower BA on ImageNet-10. }



\begin{table}[!t]
\setlength{\tabcolsep}{10pt}
\centering
    \caption{\red{Comparison (\%) of our backdoor attack with the One-to-N~\cite{xue2020one} backdoor attack on attacking four target classes simultaneously. \red{The highest BA and ASR scores are highlighted in bold.}}}
    \label{table:compareone2n}
    \begin{tabular}{llcccc} 
    \toprule
    \multirow{2}{*}{Datasets} & 
                     \multicolumn{2}{c}{\red{One-to-N~\cite{xue2020one}}}  & \multicolumn{2}{c}{Ours}\\ 
     \cmidrule(l){2-3}  \cmidrule(l){4-5}       
     & \red{BA} & \red{ASR} &BA& ASR  \\ 
     \midrule
    MNIST          & 99.26 & 99.61 & \textbf{99.36} & \textbf{99.93}  \\
    CIFAR-10 & \red{90.76} & \red{87.35} & \textbf{94.38}& \textbf{99.70}\\ 
    GTSRB    & \red{97.50} & \red{76.34} & \textbf{99.36} & \textbf{99.89}  \\ 
    ImageNet-10  & \textbf{\red{98.60}}  & \red{88.20} & 98.20 & \textbf{ 94.20}  \\
    \bottomrule
    \end{tabular}
    \vspace{-4mm}
\end{table}

\subsection{Invisibility of Triggers}
\label{stealthiness}
We test the invisibility of the triggers in our $M$-to-$N$ backdoor attack from both visual and quantitative perspectives.

\begin{table}[!t]
\setlength{\tabcolsep}{10pt}

\centering
        \caption{PSNR, SSIM and LPIPS1 (LPIPS\_VGG) and LPIPS2 (LPIPS\_Alex) average values of our $M$-to-$N$ backdoor attack.}
    \label{tab:ssim}
    \begin{tabular}{ccccc} 
        \toprule
        \#Targets & PSNR & SSIM  & LPIPS1 & LPIPS2    \\ 
        \hline
          1            &  45.87  &    0.9953  &  0.0001 &   0.0026  \\
                                  2           &  41.29  &    0.9864  &  0.0006 &   0.0110  \\
                                 4            &  44.88  &    0.9936 &  0.0041 &   0.0002  \\
                                 5            &  45.88     & 0.9951  &  0.0002   & 0.0030    \\ 
        \bottomrule
    \end{tabular}
    \vspace{-3mm}
\end{table}

\subsubsection{Visual Effects}
We conduct two different attack paradigms, namely the 1-to-2 attack ($M=1$, $N=2$) and the 2-to-1 attack ($M=2$, $N=1$), to demonstrate the visual effects of the poisoned images on ImageNet-10, as displayed in fig.~\ref{fig:visual}.
The poisoned images are indistinguishable from their clean counterparts, as shown in Figs.~\ref{fig:visual} (a) and (b). Besides, figs.~\ref{fig:visual} (c) and (g) suggest that our trigger embedding network $\mathcal{H}$ can effectively hide the triggers in the clean images. \red{The recovered triggers in Figs.~\ref{fig:visual} (e) and (i) preserve their original contents compared to the actual triggers in Figs.~\ref{fig:visual} (d) and (h).}

\subsubsection{Quantitative Results}
We evaluate the average PSNR, SSIM, and LPIPS values on the poisoned testing datasets and simply display the results of our 1-to-$N$ attack paradigm ($M=1$) on the CIFAR-10 dataset in Table~\ref{tab:ssim}. Similar results are observed on other datasets and the 10-to-$N$ attack paradigm. \red{As shown in Table~\ref{tab:ssim}, the PSNR values exceed 40 dB, SSIM values are close to 1, and LPIPS values are near 0, indicating a high degree of similarity between the poisoned and clean images. Furthermore, the SSIM and PSNR values do not significantly decrease as the number of target classes increases, suggesting the invisibility of our triggers across multiple target classes.}

\begin{figure}[!t]
	\centering
	\begin{minipage}[b]{0.99\linewidth}
		\begin{minipage}[b]{0.19\linewidth}
			\centerline{\small Clean}
		    \vspace{1mm}
			\centerline{\includegraphics[width=1\linewidth]{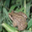}}
			\vspace{1mm}
			\centerline{\includegraphics[width=1\linewidth]{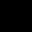}}
    		\centerline{\small PSNR (dB)}
		\end{minipage}\hfill
		\begin{minipage}[b]{0.19\linewidth}
		    \centerline{\small Input-Aware}
		    \vspace{1mm}
		    \centerline{\includegraphics[width=1\linewidth]{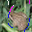}}
    		\vspace{1mm}
    		\centerline{\includegraphics[width=1\linewidth]{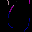}}
            \centerline{\small 19.90}
            \end{minipage}\hfill
		\begin{minipage}[b]{0.19\linewidth}
            \centerline{\small WaNet}
            \vspace{1mm}
		    \centerline{\includegraphics[width=1\linewidth]{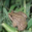}}
    		\vspace{1mm}
    		\centerline{\includegraphics[width=1\linewidth]{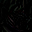}}
            \centerline{\small 28.92}
		\end{minipage}\hfill
		\begin{minipage}[b]{0.19\linewidth}
		    \centerline{\small One-to-N}
    		\vspace{1mm}
    		\centerline{\includegraphics[width=1\linewidth]{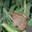}}
    		\vspace{1mm}
    		\centerline{\includegraphics[width=1\linewidth]{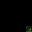}}
    		\centerline{\small 30.19}
    	\end{minipage}\hfill
    	\begin{minipage}[b]{0.19\linewidth}
		   \centerline{\small Ours}
            \vspace{1mm}
		    \centerline{\includegraphics[width=1\linewidth]{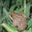}}
    		\vspace{1mm}
    		\centerline{\includegraphics[width=1\linewidth]{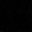}}
            \centerline{\textbf{44.54}}
		\end{minipage}\hfill
	\end{minipage}
 \caption{Generated poisoned images. The top row shows the poisoned images, while the bottom row indicates the residuals between the poisoned and clean images. }
 \label{fig: visualandpsnr}
\vspace{-4mm}
\end{figure}

\subsubsection{Comparisons with Prior Backdoor Attacks}
\label{sec: stealthcomp}
We first calculate the average PSNR values of the poisoned images in the whole poisoned testing dataset with their clean images. Note that the One-to-N and our $M$-to-$N$ attacks can simultaneously attack multiple target classes, and we set the number of the target classes as four ($N=4$) in our experiment. The average PSNR values of our $M$-to-$N$, WaNet~\cite{nguyen2021wanet}, Input-Aware~\cite{nguyen2020input} and One-to-N~\cite{xue2020one} attacks are 44.88 dB, 23.32 dB, 20.66 dB, and 29.80 dB, respectively. Our $M$-to-$N$ backdoor attack can generate poisoned images with a much larger average PSNR value than other attacks. 

We also show the visual effects of poisoning the clean image \textit{frog} for different single-target attacks and the poisoned image with the maximum PSNR value for the One-to-N attack and the poisoned image with the minimum PSNR value for our $M$-to-$N$ attack. As can be seen from Fig.~\ref{fig: visualandpsnr}, the Input-Aware~\cite{nguyen2020input} and One-to-N~\cite{xue2020one} attacks have clear triggers. Both WaNet~\cite{nguyen2021wanet} and our $M$-to-$N$ attacks can generate poisoned images indistinguishable from the originals; however, the residuals indicate that the modification to the clean image in our $M$-to-$N$ is undetectable. This result verifies the high invisibility of the triggers in our $M$-to-$N$ attack.

\subsection{Attack Robustness}
We evaluate the robustness of our $M$-to-$N$ backdoor attack under different pre-processing operations and compare it with prior backdoor attacks. These pre-processing operations include flipping, random rotation (Rotation), padding to the original size after shrinking (Shrinking\&Padding), resizing to the original size after random cropping (Cropping\&Resizing), and Gaussian noise blurring.

\red{As shown in Table~\ref{tab:robustness}, our $M$-to-$N$ attack achieves the highest ASRs against most pre-processing operations. In contrast, baseline attacks exhibit significant reductions in ASRs under similar conditions, particularly with operations such as S\&P. This vulnerability in baseline attacks could be attributed that their triggers typically modify only a small portion of poisoned images, rendering them sensitive to the change of position and appearance by preprocessing operations, as discussed in~\cite{LI20211,zhang2021poison}. The ASR of our attack under JPEG compression significantly outperforms advanced attacks such as the Input-aware and WaNet, demonstrating our attack's enhanced robustness to JPEG compression. Our attack is more susceptible to high levels of JPEG compression (\textit{i.e.}, low-quality factor) compared to the One-to-N method. This is because we use subtle perturbations as real triggers, whereas the One-to-N method affects only small pixel areas. As a result, our method introduces more high-frequency information, making it more sensitive to compression. However, the subtle nature of our triggers also enhances our attack’s robustness against other preprocessing operations like rotation or cropping, making it resilient in various scenarios. }

\begin{figure*}[!t]
	\centering
	\begin{minipage}[b]{0.95\linewidth}
    	\begin{minipage}[b]{0.24\linewidth}
			\centerline{\includegraphics[width=1\linewidth]{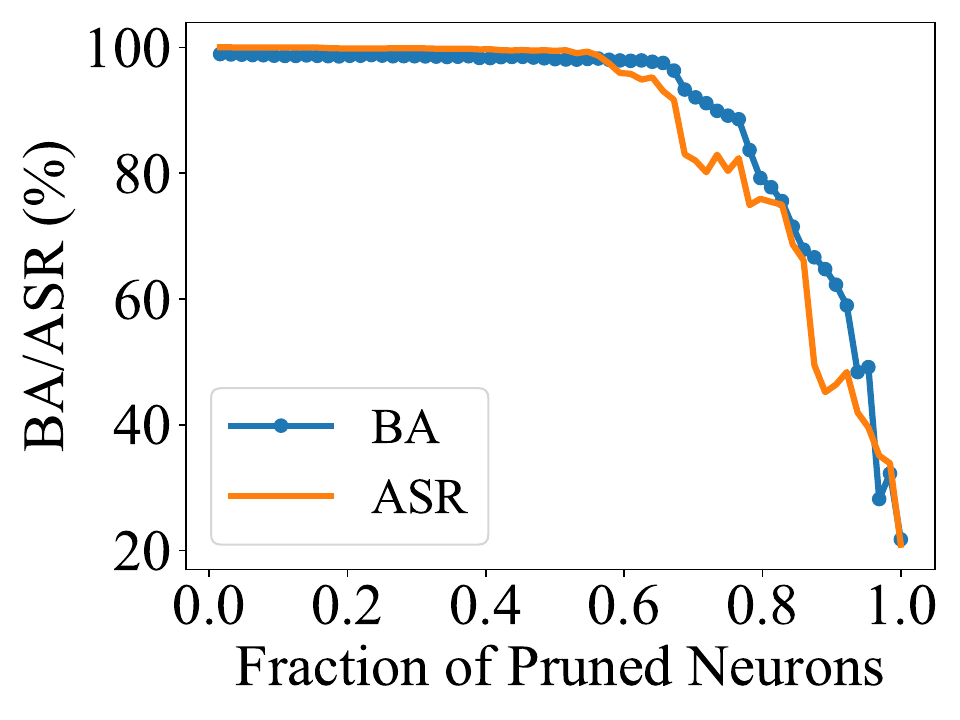}}
   \centerline{(a)}
		\end{minipage}	
            \begin{minipage}[b]{0.24\linewidth}
			\centerline{\includegraphics[width=1\linewidth]{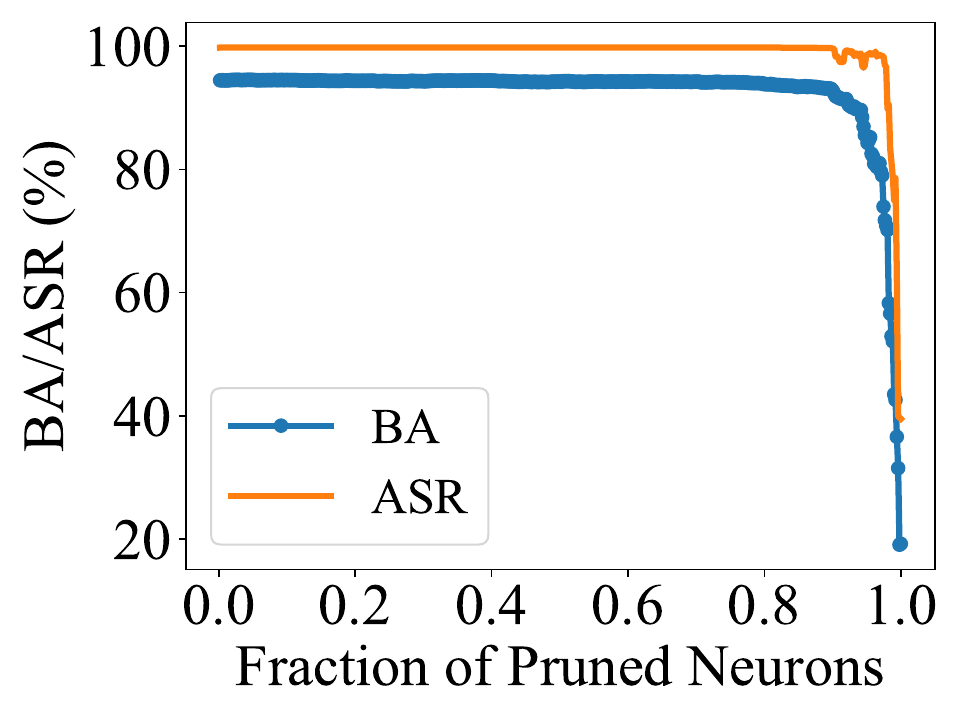}}
            \centerline{(b)}
		\end{minipage}
            \begin{minipage}[b]{0.24\linewidth}
			\centerline{\includegraphics[width=1\linewidth]{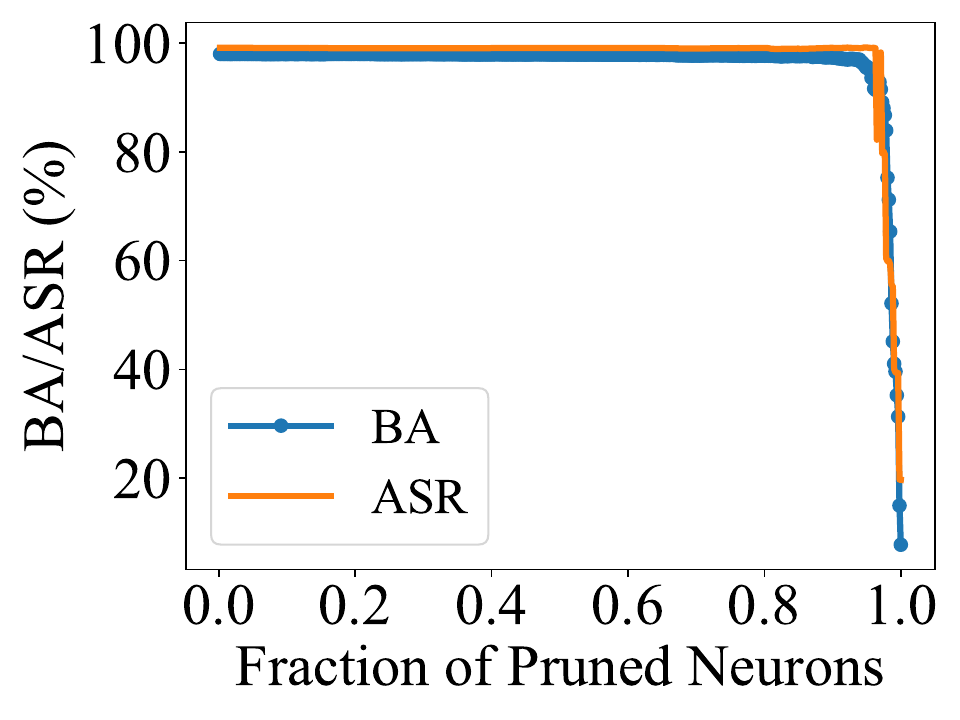}}
   \centerline{(c)}
		\end{minipage}
            \begin{minipage}[b]{0.24\linewidth}
			\centerline{\includegraphics[width=1\linewidth]{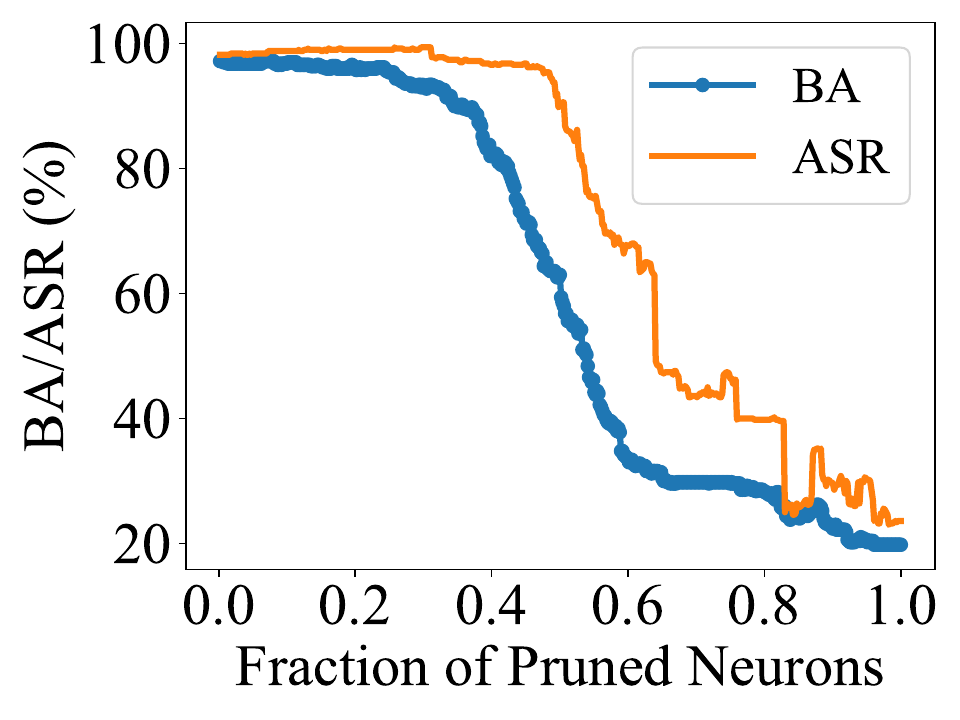}}
   \centerline{(d)}
		\end{minipage}
	\end{minipage}
	\caption{Ability of our $M$-to-$N$ backdoor attack to resist the fine-pruning defence on different datasets. Results with trigger number $M=1$ for (a) MNIST, (b) CIFAR-10, (c) GTSRB, and (d) ImageNet-10 datasets when simultaneously attacking five target classes.}
 \label{fig:fine-strip}
 \vspace{-3mm}
\end{figure*}

\begin{table}[!t]
\centering
\caption{Robustness of different backdoor attacks against common pre-processing operations. The values are the ASRs (\%) on the CIFAR-10 dataset (The ASRs of the One-to-N attack and our 1-to-4 attack are obtained on testing datasets poisoned by four triggers equally.). \red{The highest ASR scores for single-target and multi-target attacks are in bold.}}
\label{tab:robustness}
\begin{threeparttable}
\small
\begin{tabular}{lcccccc} 
\toprule
\multirow{2}{*}{ OPs.} & \multicolumn{3}{c}{Single-target} & \multicolumn{2}{c}{Multi-target}  \\ \cmidrule(lr){2-4}\cmidrule(l){5-6}
 
& Input-Aware & WaNet & $1$-to-$1$ & One-to-N & $1$-to-$4$ \\
\midrule
 None &99.32 & \textbf{99.86} & 99.76 & 87.35 & \textbf{99.70} \\
Flipping & 74.26 & 76.29 & \textbf{97.18} & \red{57.87} & \textbf{96.49} \\
Rotation & 75.71 & 17.17 & \textbf{95.30} & \red{51.68} & \textbf{95.11} \\
S\&P\tnote{1} & 78.60 & \textbf{99.69} & 93.98 & \red{12.91} & \textbf{91.59} \\
C\&R\tnote{2} & 81.65 & \textbf{99.01} & 96.51 & \red{87.62} & \textbf{96.35} \\
GBL\tnote{3}  & 70.57 & 12.22 & \textbf{85.00} & \red{80.63}& \textbf{90.27} \\
 \red{JPEG\_50}&\red{ 1.80} & \red{0.48}&\red{\textbf{44.77}} &\red{\textbf{57.71}}& \red{44.07}\\
 \red{JPEG\_75} & \red{1.67}& \red{0.55}& \red{\textbf{47.37}} &\red{\textbf{68.33}} & \red{47.65}\\
\bottomrule
\end{tabular}
    \begin{tablenotes}   
        \footnotesize 
        \item[1] ``S\&P'' indicates Shrinking\&Padding.
        \item[2] ``C\&R'' means Cropping\&Resizing.
        \item[3] ``GBL'' means Gaussian noise blurring.         
      \end{tablenotes}   
\end{threeparttable}
\vspace{-3mm}
\end{table}

\section{Evading Existing Defenses}
\label{sec: Resdefences}
\red{We evaluate the ability of our defense to evade existing defenses, including Fine-Pruning~\cite{liu2018fine}, Neural Cleanse~\cite{wang2019neural}, Neural Attention Distillation~\cite{li2021neural} and Adversarial Neuron Pruning~\cite{wu2021adversarial}, Meta Neural Trojan Detection~\cite{xu2021detecting}, STRIP~\cite{gao2019strip}, and SentiNet~\cite{chou2020sentinet}. Note that our attack exhibits similar resistance against these defenses with different settings of $M$ and $N$, and thus we mainly display the results of our $1$-to-$N$ ($M=1$) attack. The results of our $10$-to-$N$ ($M=10$) attack are shown in the supplementary material.}


\subsection{Fine-Pruning}
As discussed in Section~\ref{sec: defence}, the fine-pruning (FP) defense directly removes backdoors by pruning the dormant neurons when inputting the clean images.
Following the settings in previous methods\cite{nguyen2021wanet,nguyen2020input,li2021invisible}, we also evaluate the resistance of our $M$-to-$N$ attack to FP by analyzing the activation of the neurons in the last convolutional layer. 
Fig.~\ref{fig:fine-strip} displays the pruning effects of the $1$-to-$N$ attack when attacking five target classes simultaneously (i.e., $M=1,N=5$) on the four datasets. It can be seen that the ASRs of our attack only decrease slightly with the increase of the fraction of pruned neurons. 
Furthermore, the ASRs of our attack decrease after the BAs, indicating that FP cannot remove our backdoors without impacting the performance of original tasks.

\begin{figure}[!t]
	\begin{minipage}[b]{0.99\linewidth}
		\begin{minipage}[b]{0.49\linewidth}
			\centerline{\includegraphics[width=1\linewidth]{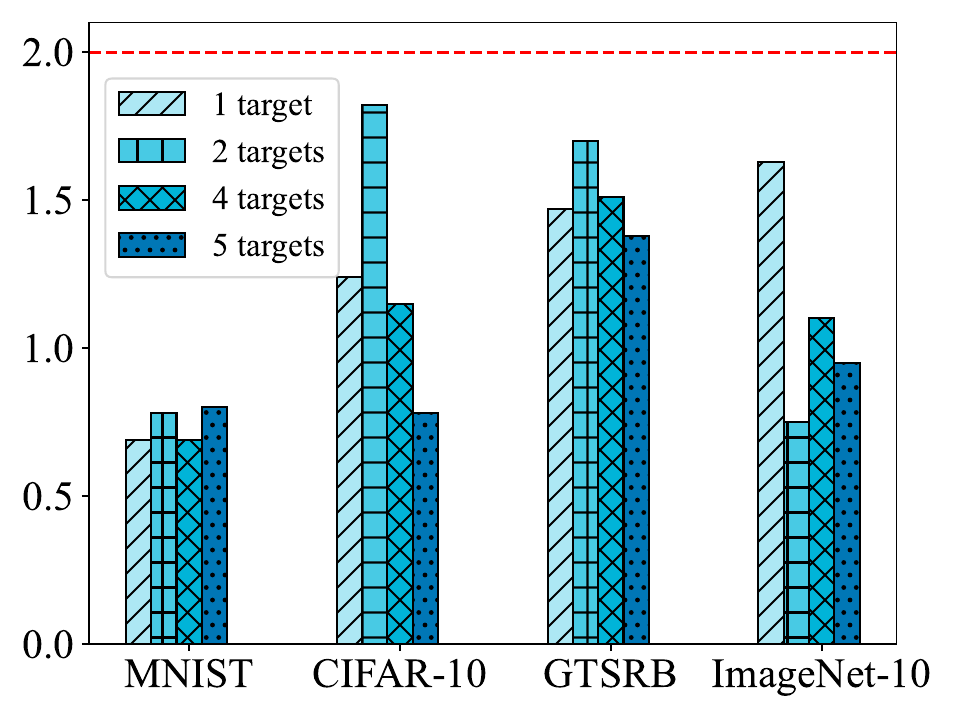}}
                \vspace{-2mm}
			\centerline{(a)}
		\end{minipage}\hfill
		\begin{minipage}[b]{0.49\linewidth}
			\centerline{\includegraphics[width=1\linewidth]{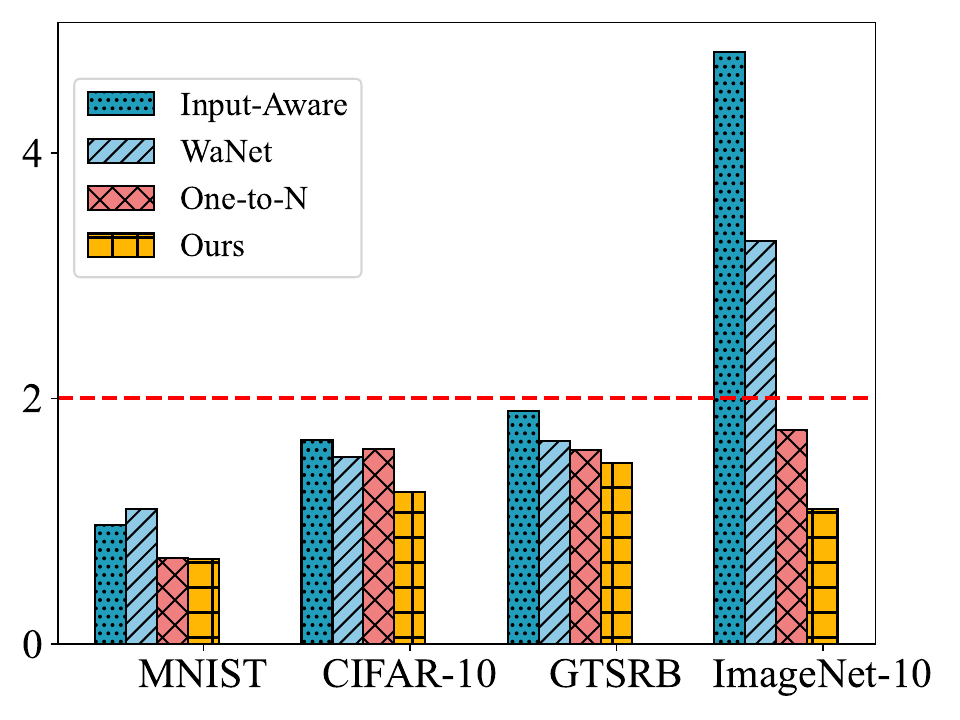}}
            \vspace{-2mm}			
            \centerline{(b)}
		\end{minipage}
	\end{minipage}
	\caption{\red{Ability of our $M$-to-$N$ backdoor attack to resist NC. The anomaly indexes of our attack under case (a) a single trigger ($M=1$) per target class, and case. (b) Anomaly index comparison of different backdoor attacks with our 1-to-4 attack ($M=1$ and $N=4$).}}
	\label{fig:nc}
 \vspace{-3mm}
 \end{figure}

 \begin{figure}[!t]
    \begin{minipage}[b]{0.99\linewidth}
        \begin{minipage}[b]{0.49\linewidth}
            \centerline{\includegraphics[width=1\linewidth]{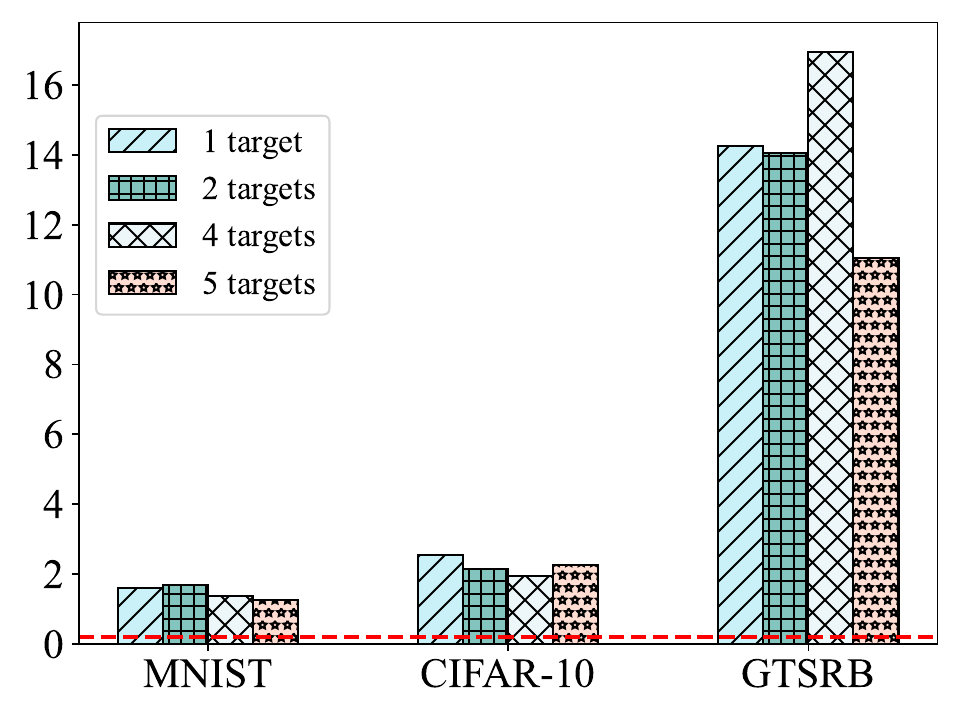}}
            \centerline{(a)}
        \end{minipage}\hfill
        \begin{minipage}[b]{0.49\linewidth}
            \centerline{\includegraphics[width=1\linewidth]{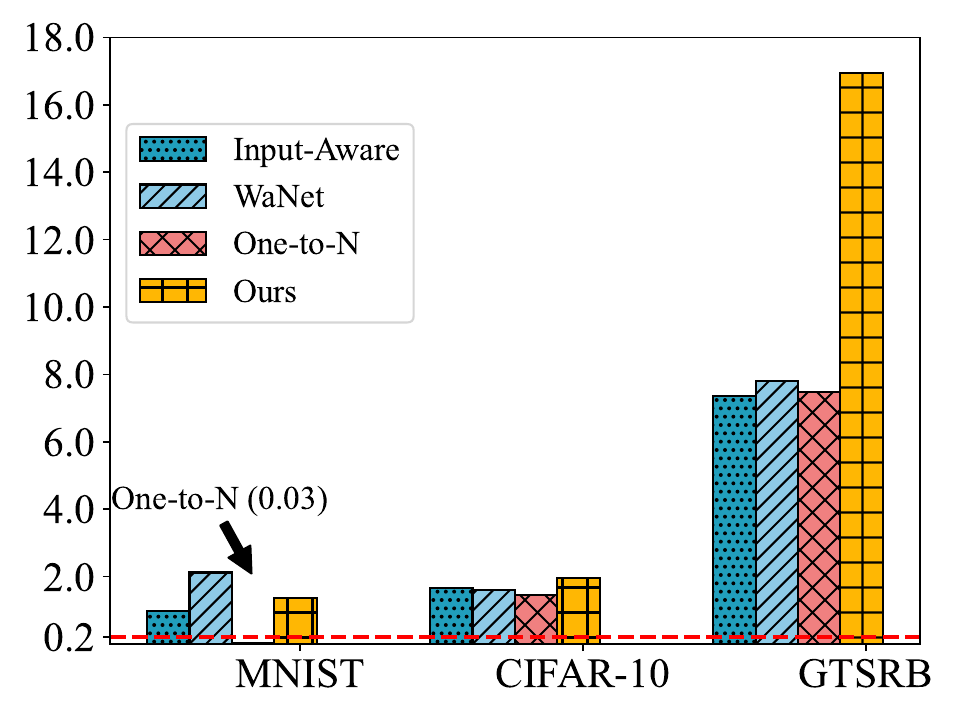}}
            \centerline{(b)}
        \end{minipage}\hfill
    \end{minipage}
    \caption{\red{Ability of our $M$-to-$N$ backdoor attack to resist the STRIP defense. (a) The minimum entropy of our attack with a single trigger ($M=1$) for different targets. (b) The comparison of the minimum entropy among different attacks with our 1-to-4 attack ($M=1$ and $N=4$).}}
	\label{fig: stripgrid}
\end{figure} 

\begin{table}[!t]
\centering
\caption{Performance (\%) of our $M$-to-$N$ backdoor attack to resist the ANP and NAD defenses on the MNIST, CIFAR-10, and GTSRB datasets when attacking five target classes.}
\label{table: ANP}
\begin{tabular}{cccccc}         
\toprule
\multirow{2}{*}{Dataset} & \multirow{2}{*}{Defence} &\multicolumn{2}{c}{$M=1$} & \multicolumn{2}{c}{$M=10$}\\
\cmidrule(l){3-4} \cmidrule(l){5-6} 
& & BA& ASR & BA & ASR\\
\midrule
\multirow{3}{*}{MNIST} &  None &  99.1 &  99.81&  99.27&  98.51\\
&  ANP &  6.4 &  19.75 &  11.35 &  20.32\\
&  NAD&  11.35 &  19.87 &  11.35&  20.32 \\
\hline   
\multirow{3}{*}{CIFAR-10}&  None &  93.32&  99.72 &  93.87 &  99.81\\
&  ANP&  93.02 &  88.93 &  93.09&  90.93\\
&  NAD &  40.78 &  10.88 &  36.88 &  7.66\\
\hline 
\multirow{3}{*}{GTSRB} &  None&  99.22&  99.64 &  99.35&  99.89 \\
&  ANP &  94.85 &  98.91 &  96.18 &  98.36 \\
&  NAD &  7.11 &  0.18 &  8.31&  1.26 \\
\bottomrule
\end{tabular}
\end{table}

\subsection{Neural Cleanse}
\red{NC reverses a trigger for each class label by transforming the predictions of all poisoned images to the label and applying anomaly detection to discriminate the triggers that are notably smaller than the rest. The anomaly index threshold is set at two, and any model with an anomaly index larger than two is considered a backdoored model. The results in Figs.~\ref{fig:nc}(a) reveal that the average anomaly indices of our attack consistently remain below this threshold. Also, Fig.~\ref{fig:nc}(b) illustrates that our attack achieves the lowest anomaly index values compared to other attacks, exhibiting higher robustness.}

\subsection{STRIP}
\begin{figure*}[!t]
	\centering
	\begin{minipage}[b]{0.95\linewidth}
    	\begin{minipage}[b]{0.245\linewidth}
			\centerline{\includegraphics[width=1\linewidth]{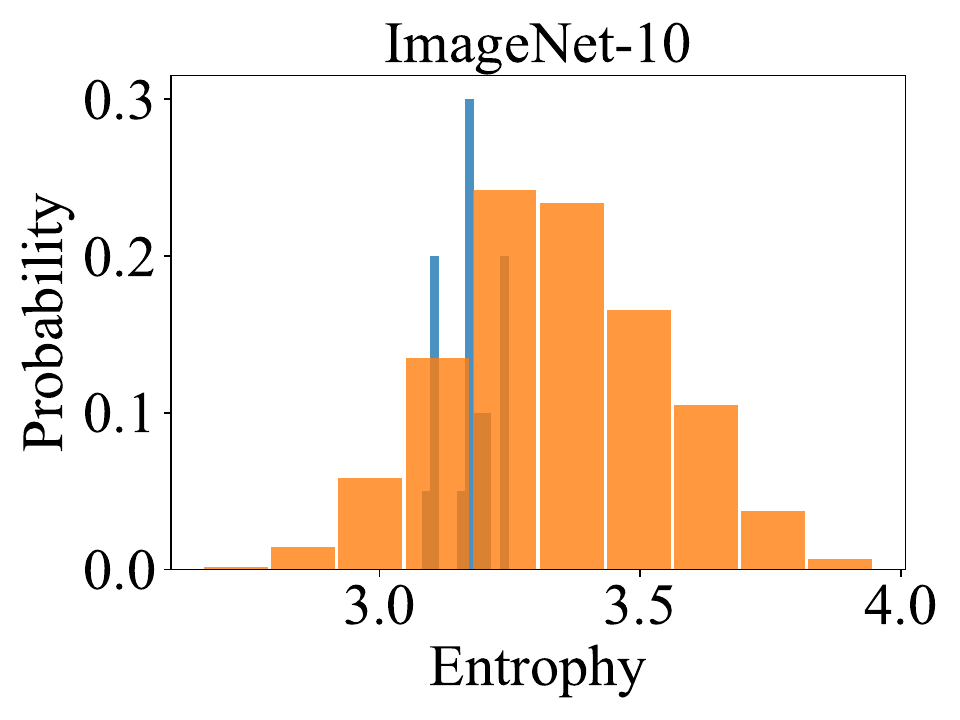}}
    \vspace{-2mm}
			\centerline{(a)}
		\end{minipage}	
            \begin{minipage}[b]{0.245\linewidth}
			\centerline{\includegraphics[width=1\linewidth]{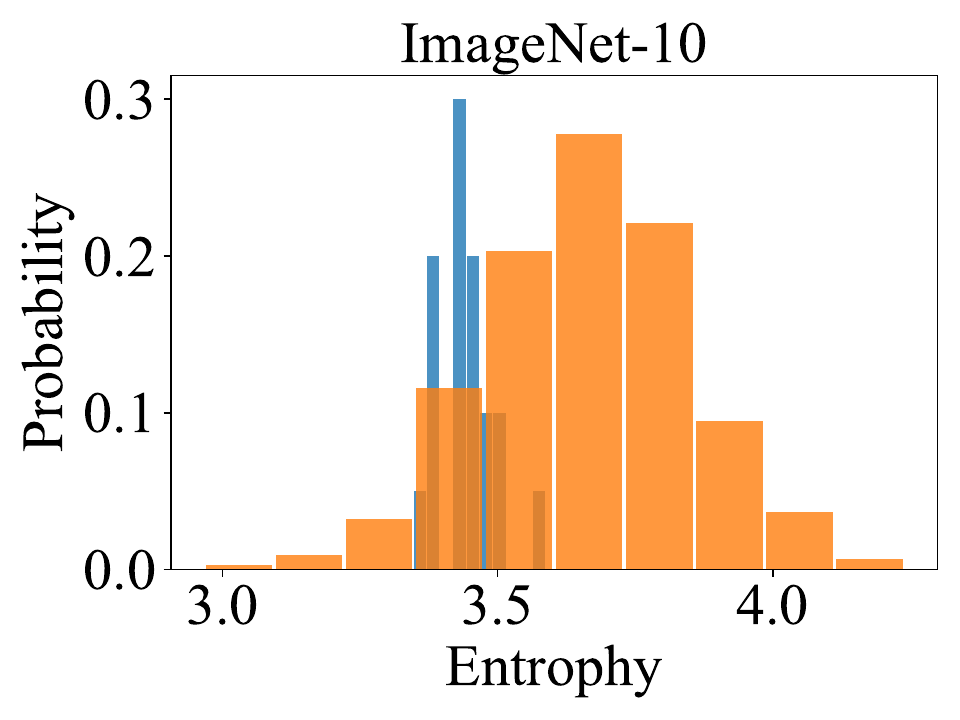}}
    \vspace{-2mm}
			\centerline{(b)}
		\end{minipage}
            \begin{minipage}[b]{0.245\linewidth}
			\centerline{\includegraphics[width=1\linewidth]{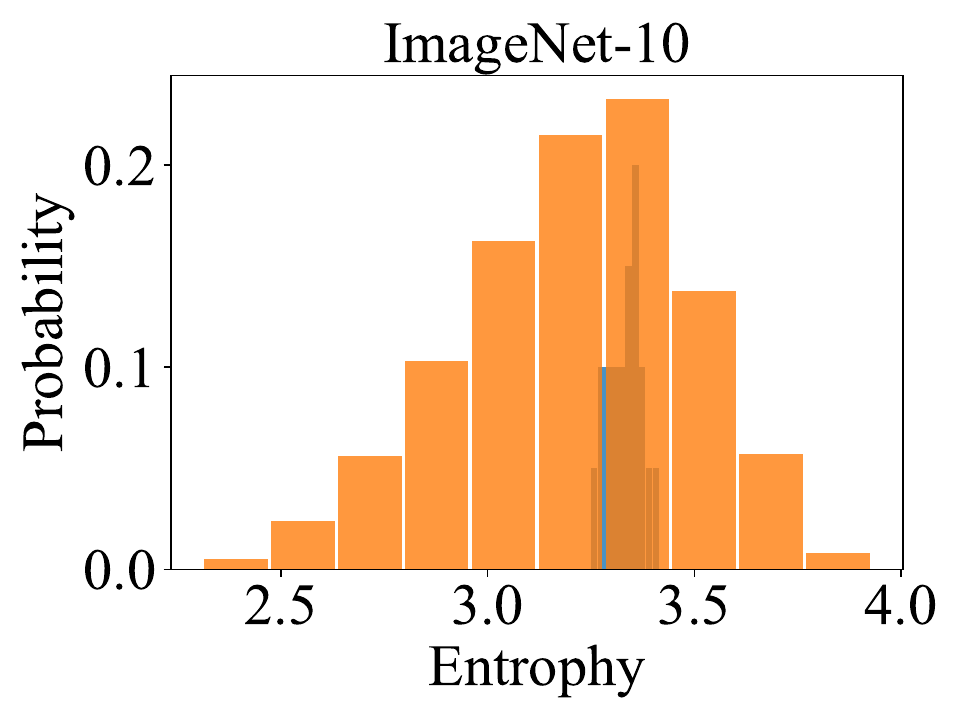}}
    \vspace{-2mm}
			\centerline{(c)}
		\end{minipage}
            \begin{minipage}[b]{0.245\linewidth}
			\centerline{\includegraphics[width=1\linewidth]{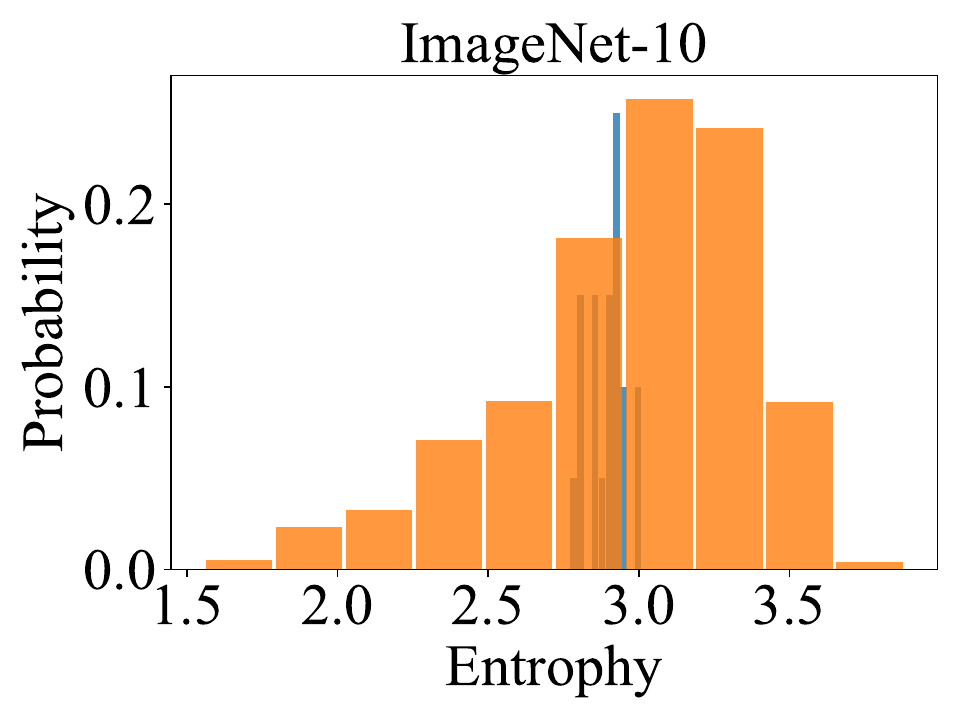}}
    \vspace{-2mm}
			\centerline{(d)}
		\end{minipage}
	\end{minipage}
 \vspace{-2mm}
	\caption{Ability of our $M$-to-$N$ backdoor attack to resist the STRIP defence on the ImageNet-10 dataset. (a)-(d) represent results of different values of $M$ and $N$. (a) 1-to-1 attack ($M=1$, $N=1$), (b) 2-to-1 attack, (c) 1-to-5 attack, (d) 2-to-5 attack.}
    \label{fig: strip-imagenet}
 \vspace{-2mm}
\end{figure*}

\red{STRIP~\cite{gao2019strip} perturbs inputs with clean images and calculates the entropy of the resulting mixture predictions. A low entropy indicates a violation of the input-dependence property of a clean model and suggested a backdoor image. The entropy boundary is set at 0.2, and any image with an entropy below this value is considered a backdoor image. Fig.~\ref{fig: stripgrid}(a) shows that all entropy values of our attack exceed the detection threshold, and 
Fig.~\ref{fig: stripgrid}(b) shows that our $M$-to-$N$ attack achieves a larger minimum entropy than other attacks in most cases. The results indicate the strong ability of our attack to withstand the STRIP defense. However, One-to-N exhibits lower entropy on MNIST due to its strategy of modifying edge pixel values, which are uniformly consistent across MNIST images, resembling a variant of BadNets. Additionally, Fig.~\ref{fig: strip-imagenet} illustrates consistent entropy distributions for clean and backdoor images on the ImageNet-10 dataset, across various target classes and triggers.}

\subsection{Two Post-Training-Based Defenses}
We also evaluate the robustness of our $M$-to-$N$ attack against two advanced model-based defenses, Neural Attention Distillation (NAD)~\cite{li2021neural} and Adversarial Neuron Pruning (ANP)~\cite{wu2021adversarial}. ANP applies fine-pruning to remove the neurons sensitive to adversarial perturbations, while NAD leverages a teacher model to guide the fine-tuning process of backdoored DNN models. Following the settings of NAD and ANP, we assume that defenders have a subset of clean images from the original training dataset (5\% for NAD and 1\% for ANP), and evaluate the effectiveness of these defenses against our attack with different numbers of triggers. Table~\ref{table: ANP} shows the BAs and ASRs of our backdoored models targeting five target classes simultaneously ($N=5$) on different datasets after undergoing the backdoor erasing processes of NAD and ANP. As we can see, the ANP defense has minimal impact on the BAs and ASRs of the $M$-to-$N$ attack on the CIFAR-10 and GTSRB datasets, while significantly reducing both metrics on the MNIST datasets. The results indicate that our attack is resistant to the ANP defense. For NAD, while ASRs decrease, BAs degrade even more significantly, demonstrating the resistance of our attack to the NAD defense.
\begin{figure}[!t]
             \centering
            \includegraphics[width=0.3\textwidth]{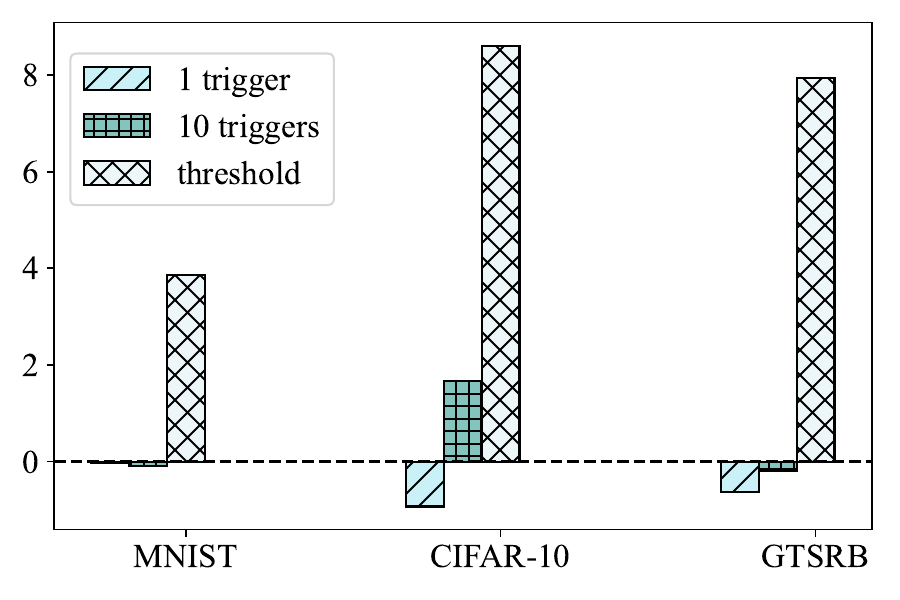}
            \caption{Ability of our $M$-to-$N$ attack to resist the MNTD defense on different datasets when attacking five target classes \protect\footnotemark[1].}
            \vspace{-2mm}
            \label{fig: MNTD}
\end{figure}
 \stepcounter{footnote} 
 \footnotetext{The meta-classifier's score is not constrained by the sigmoid activation function and does not directly indicate the probability of a backdoor in the model. A DNN model is classified as backdoored if its score exceeds the threshold; otherwise, it is considered benign.}

\subsection{Meta Neural Trojan Detection}
Meta Neural Trojan Detection (MNTD)~\cite{xu2021detecting} trains a meta-classifier to predict whether a given model is a backdoored model or not. The defender first calculates a threshold on the training dataset, and a suspicious model is considered backdoored if the prediction score exceeded the threshold. We test the resistance of our $M$-to-$N$ attack against the MNTD defense on the MNIST, CIFAR-10, and GTSRB datasets. Fig.~\ref{fig: MNTD} shows the prediction scores of the trained meta-classifier on the $M$-to-$N$ attack targeting five target classes simultaneously ($N=5$) with different numbers of triggers $M~(M=1,10)$. As can be seen, all the prediction scores on the three datasets are smaller than their respective thresholds, demonstrating that our $M$-to-$N$ attack is resistant to MNTD.

 \begin{figure}
		\begin{minipage}[b]{0.19\linewidth}
                \centerline{\includegraphics[width=1\linewidth]{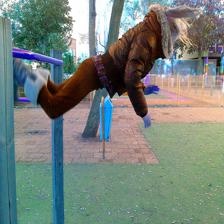}}
                \vspace{1mm}
                \centerline{\includegraphics[width=1\linewidth]{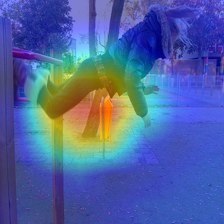}}
                \centerline{ (a)}
		\end{minipage}
		\begin{minipage}[b]{0.19\linewidth}
                \centerline{\includegraphics[width=1\linewidth]{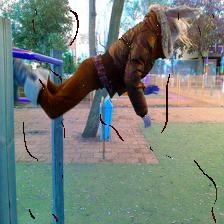}}
                \vspace{1mm}
                \centerline{\includegraphics[width=1\linewidth]{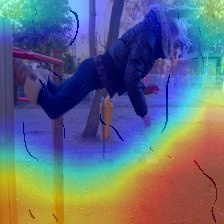}}
			\centerline{ (b)}
		\end{minipage}
		\begin{minipage}[b]{0.19\linewidth}
                \centerline{\includegraphics[width=1\linewidth]{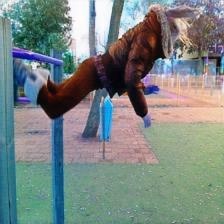}}
                \vspace{1mm}
                \centerline{\includegraphics[width=1\linewidth]{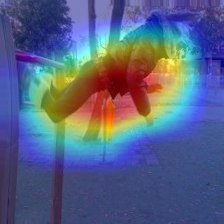}}
			\centerline{ (c)}
		\end{minipage}
		\begin{minipage}[b]{0.19\linewidth}
            \centerline{\includegraphics[width=1\linewidth]{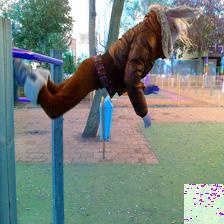}}
            \vspace{1mm}
            \centerline{\includegraphics[width=1\linewidth]{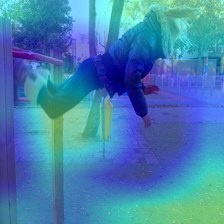}}
            \centerline{ (d)}
		\end{minipage}
		\begin{minipage}[b]{0.19\linewidth}
                \centerline{\includegraphics[width=1\linewidth]{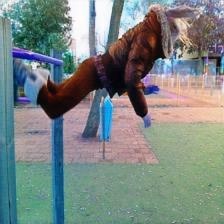}}
                \vspace{1mm}
                \centerline{\includegraphics[width=1\linewidth]{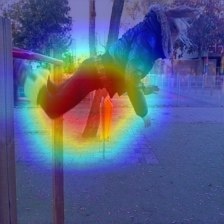}}
        	\centerline{ (e)}
		\end{minipage}
   \vspace{-5mm}
   \caption{The poisoned images generated by various attacks, along with their corresponding saliency maps, highlight the most important regions in the images.} (a) Clean images; (b) Input-Aware~\cite{nguyen2020input}; (c) WaNet~\cite{nguyen2021wanet}; (d) One-to-N~\cite{xue2020one}; (d) Ours. 
 \label{fig:gradcam}
  \vspace{-3mm}
\end{figure}

\subsection{SentiNet}
The SentiNet~\cite{chou2020sentinet} defense utilized the input saliency maps to detect potential trigger regions within incoming inputs. The results in Fig.~\ref{fig:gradcam} show that our poisoned images consistently exhibit substantial areas of overlap with the corresponding clean images, and the trigger regions of the baseline defenses are readily identifiable. This is because our triggers are related to the target classes and are well distributed in the clean images via the poisoned image generation framework. The experimental results demonstrate that our $M$-to-$N$ attack can withstand the SentiNet defense.


\section{Discussion}
\label{discussion}
\subsection{Effects of the Poisoning Ratio}
\begin{figure}
        \hfill
        \begin{minipage}[b]{0.49\linewidth}
        			\centerline{\includegraphics[width=1.05\linewidth]{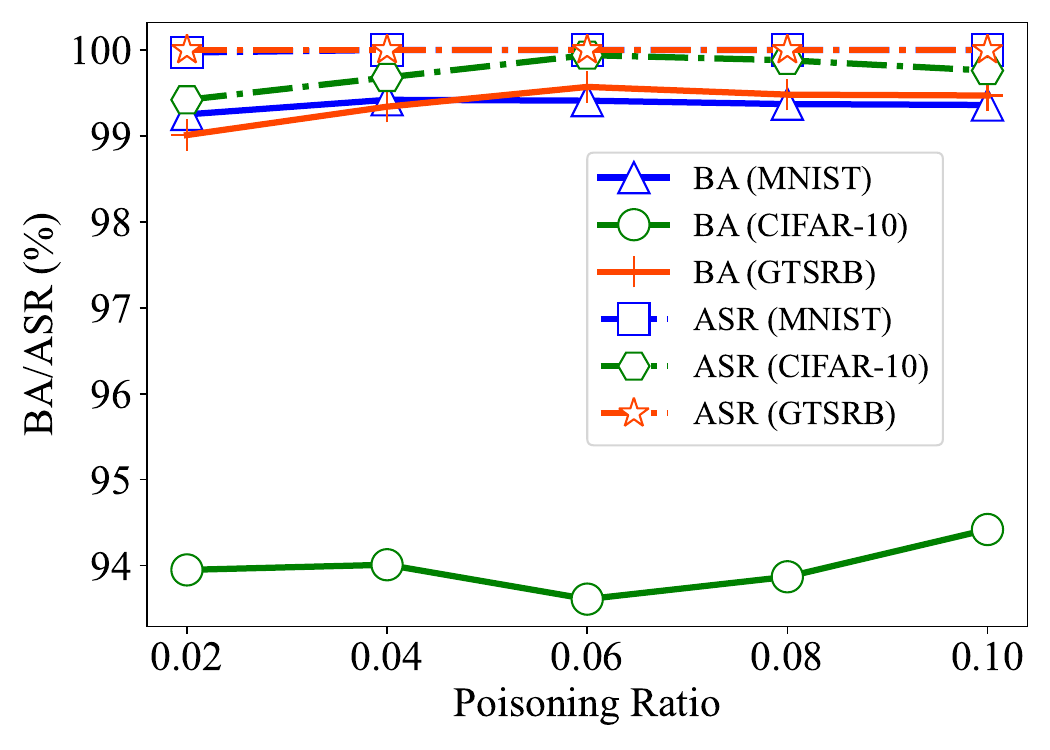}}
        			\centerline{(a)}
           \end{minipage}
           \begin{minipage}[b]{0.49\linewidth}
        			\centerline{\includegraphics[width=1.05\linewidth]{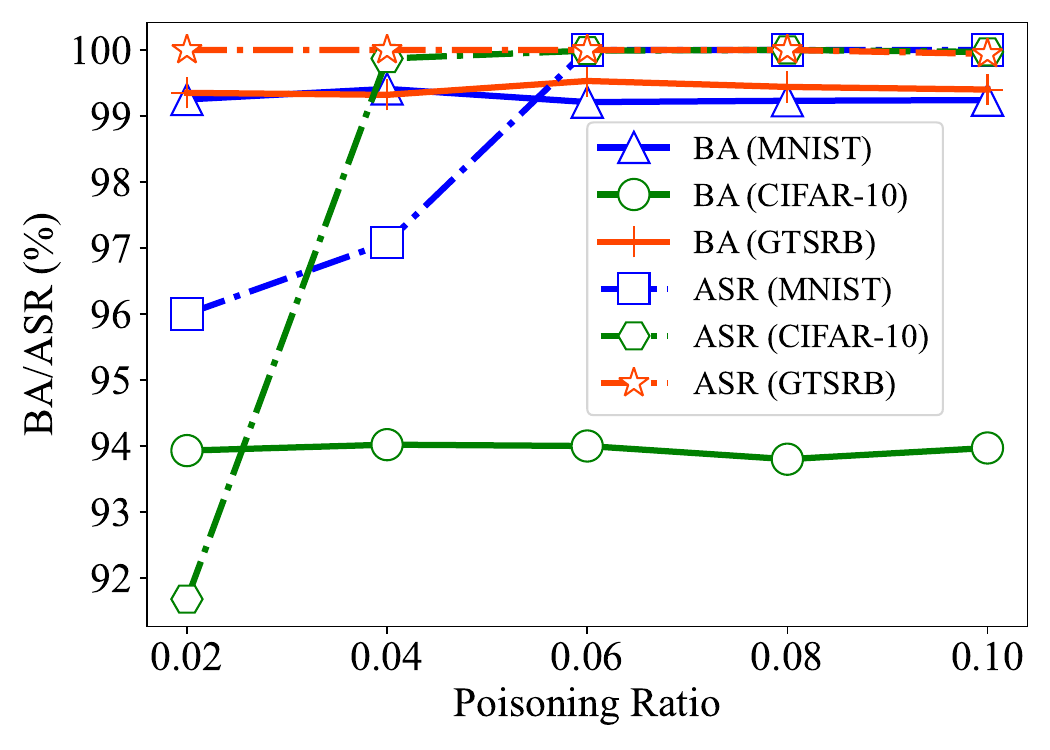}}
        			\centerline{(b)}
            \end{minipage}
            \caption{Effects of the poisoning ratio ($\rho$) to our $M$-to-$N$ backdoor attack on the three datasets. (a) 1-to-1 attack ($M=1$, $N=1$); (b) 10-to-1 attack ($M=10$, $N=1$). }
        \label{fig: poisonrate}
        \vspace{-2mm}
\end{figure}

We test the BAs and ASRs of the 1-to-1 ($M=1$, $N=1$) and 10-to-1 ($M=10$, $N=1$) attacks on different datasets when $\rho$ gradually increases from 2\% to 10\%. The results on the three datasets are shown in Fig.~\ref{fig: poisonrate}. As can be seen, our attacks can attain high ASRs (more than 95\%) by poisoning only 2\% of the training dataset. 
Besides, the ASRs increase with the increase of $\rho$ and stabilize near 100\%, while the BAs remain almost unchanged. 

\subsection{Effects of Grayscale Conversion}
During the poisoned image generation process, each RGB trigger image is converted to a grayscale image before concatenating with a clean image. Converting trigger images into grayscale format can simplify the input space of the trigger embedding network $\mathcal{H}$ and reduce the computational complexity during poisoned image generation, as an RGB color image has three channels while a grayscale image has only one. Besides, color information could mislead the recovery network $\mathcal{R}$ into recovering the false triggers. To verify this, we test the effectiveness of the poisoned images using the color trigger images directly in the poisoned image generation.  We display the results of our 1-to-4 ($M=1$, $N=4$) attack on the ImageNet-10 dataset in Table~\ref{tab:gray} as high-resolution images pose a greater challenge in hiding and recovering. As we can see, the ASR is only 52.7\%, significantly lower than those using grayscale versions of the triggers, such as 94.2\%.

\begin{table}
\setlength{\tabcolsep}{20pt}

    \centering
    \caption{Effects of trigger grayscale conversion in the poisoned image generation framework. }
    \label{tab:gray}
    \begin{tabular}{lcc} 
        \toprule
         Trigger       & BA (\%)& ASR (\%)   \\ 
        \midrule 
        Natural & 97.80  & 52.7 \\
        Grayscale & 98.2&  94.2\\
        \bottomrule
    \end{tabular}
     \vspace{-3mm}
\end{table}

\subsection{Trigger Selection}
\red{In our M-to-N attack, we randomly select clean images corresponding to the target classes to serve as triggers. Because all images in the dataset share the same distribution, different selections of clean images produce similar results. We also conduct experiments using different clean images as triggers. Specifically, for a 10-to-4 ($M$=10, $N$=4) attack on the CIFAR-10 dataset, we conducted two experiments with different randomly selected triggers. The results showed BAs of 93.9\% and 93.88\%, and ASRs of 99.87\% and 99.95\%, respectively. The variations in BAs and ASRs were within 1\%.}

\subsection{Trigger Sensitivity}
According to the discussions in~\cite{nguyen2020input,li2021invisible,lin2020composite}, the static triggers can be easily reverse-engineered and detected, whereas the dynamic triggers and those with clean features can not be detected.
In our attack, the semantic information of a trigger can be spread all over the whole clean image through the trigger embedding network $\mathcal{H}$. Thus the poisoned images are determined by the original clean image and trigger, which both consist of clean features. The residual between the poisoned and clean images is sample-specific and unique, which cannot be reverse-engineered. We also put detailed discussions in the supplementary material.

\subsection{Ablation Studies}
Our poisoned image generation framework includes the trigger embedding network $\mathcal{H}$, recovery network $\mathcal{R}$, and discriminator $\mathcal{D}$. We design ablation studies to explore the effect of each module in our framework. We implement a 1-to-4 ($M=1$, $N=4$) attack on the CIFAR-10 dataset to evaluate the effectiveness of our $M$-to-$N$ attack and the invisibility of the triggers under the three modules. 

The results in Table~\ref{tab:ablation} show that the ASR of the poisoned images generated using only the $\mathcal{H}$ module is low (11.08\%), indicating that the backdoor attack is ineffective under the settings. However, the ASR increases from 11.08\% ($\mathcal{H}$) to 99.36\% ($\mathcal{H}$ and $\mathcal{R}$) and to 99.60\% ($\mathcal{H}$, $\mathcal{R}$ and $\mathcal{D}$) when the $\mathcal{R}$ module is added into the poisoned image generation framework. This shows that the $\mathcal{R}$ module is useful for improving the ASRs. Compared to the poisoned images generated solely using the $\mathcal{H}$, the addition of the $\mathcal{R}$ leads to a decrease in the PSNR value because some trigger features are correctly incorporated into the poisoned images. However, when the $\mathcal{D}$ module is added, both the ASR and PSNR values show improvement. This proves that all three modules are essential for achieving effectiveness and invisibility of the triggers.

\begin{table}[t]
 \setlength{\tabcolsep}{10pt}
 \centering
\caption{Effects of different modules. }
\label{tab:ablation}
\begin{tabular}{lccc} 
\toprule
Modules        & BA (\%) & ASR (\%) &  PSNR (dB)    \\ 
\midrule
$\mathcal{H}$ & 88.89 & 11.08 & 49.26\\
$\mathcal{H}$ and $\mathcal{R}$ & 93.90  & 99.36 & 43.51\\
$\mathcal{H}$, $\mathcal{R}$ and $\mathcal{D}$ & 94.38 & 99.70 & 44.82\\
\bottomrule
\end{tabular}
 \vspace{-3mm}
\end{table}

\section{Conclusion}
\label{sec: conclude}
In this paper, we proposed the $M$-to-$N$ backdoor attack, a new attack paradigm to attack $N$ target classes simultaneously, and the backdoor of each target class can be activated by any one of $M$ triggers. 
To avoid introducing external features into the training dataset, we use the clean images selected from the training dataset as triggers and design a poisoned image generation framework to embed the triggers in clean images in an invisible manner. 
Extensive experimental results validate that the backdoors of multiple target classes can be effectively activated by poisoning only a small part of the training dataset. Besides, our attack has strong ability to resist pre-processing operations and state-of-the-art defenses. \red{Our future work will focus on exploring new defense strategies against the multi-target and multi-trigger paradigm, as this attack approach presents a more powerful and flexible backdoor threat to DNNs. }

\bibliographystyle{IEEEtran}
\bibliography{ref}


\begin{IEEEbiography}[{\includegraphics[width=.95\linewidth]{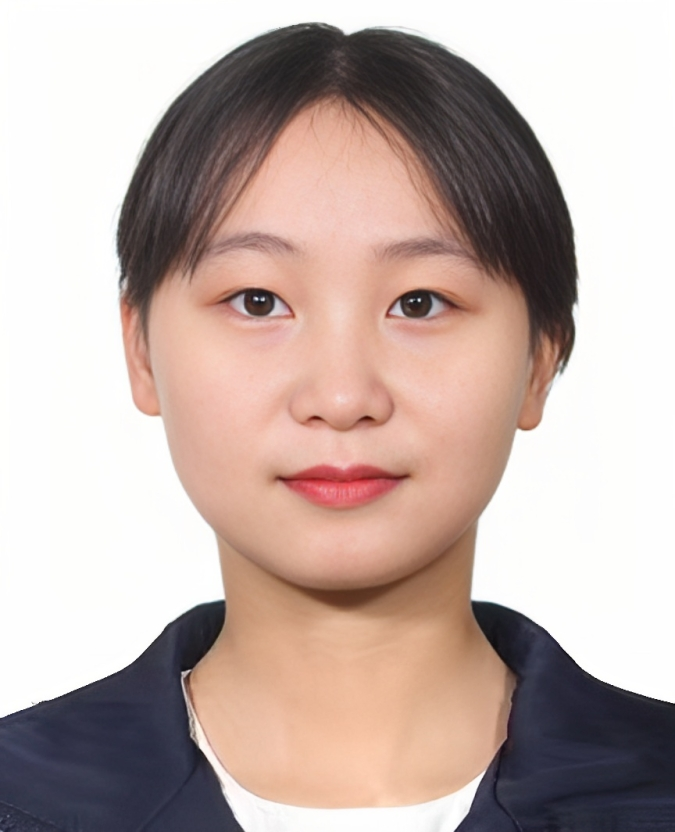}}]
    {Linshan Hou} 
    received the B.S. and M.S. degrees in Computer Science and Technology from Yanbian University, China, in 2019 and Harbin Institute of Technology, China, in 2021, respectively. She is currently pursuing the Ph.D. degree with Harbin Institute of Technology, Shenzhen, China. 
    Her research interests include AI security, especially backdoor learning and AI copyright protection.
\end{IEEEbiography}
    
\begin{IEEEbiography}[{\includegraphics[width=.95\linewidth]{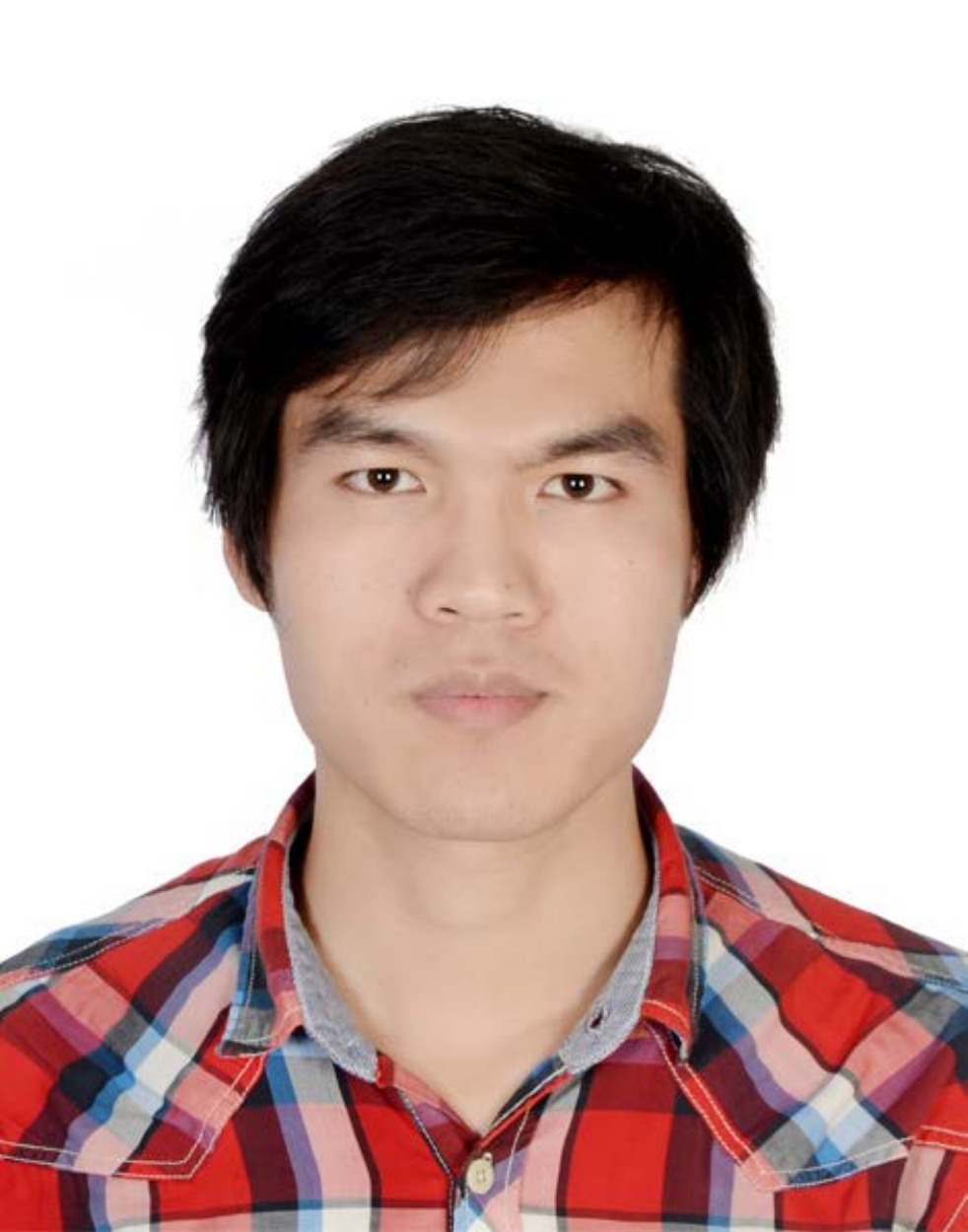}}]
    {Zhongyun Hua} (Senior Member, IEEE) 
    received the B.S. degree in software engineering from Chongqing University, Chongqing, China, in 2011, and the M.S. and Ph.D. degrees in software engineering from the University of Macau, Macau, China, in 2013 and 2016, respectively. 

   He is currently a Professor with the School of Computer Science and Technology, Harbin Institute of Technology, Shenzhen, China. His works have appeared in prestigious venues such as IEEE TDSC, IEEE TIFS, IEEE TSP, ICML, and CVPR. He has been recognized as a 'Highly Cited Researcher 2023'. His current research interests include chaotic system, multimedia security, and secure cloud computing. He has published about 100 papers on the subjects, receiving more than 7000 citations. He is an associate editor of \textit{International Journal of Bifurcation and Chaos} and \textit{IEEE Signal Processing Letters}.
\end{IEEEbiography}

\begin{IEEEbiography}[{\includegraphics[width=.95\linewidth]{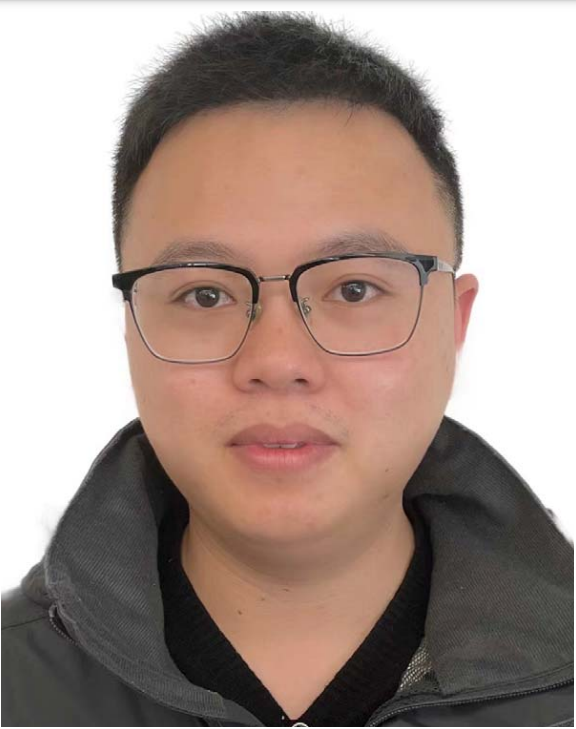}}]
    {Yuhong Li} received the bachelor’s degree in soft813 ware engineering from Xidian University, Xi’an, China, in 2010, and the Ph.D. degree from the Department of CIS, University of Macau, Macau, China, in 2016. 
    
    He is currently a Senior Staff Engineer with the Security Department, Alibaba Group, Beijing China. He has authored more than 20 papers on top-tier conferences and journals, including VLDB, ICDE, ICASSP, CIKM, IEEE TVCG, TVCG, and EDBT. His current research interests include AI for social good, multimodal understanding, self-supervised learning on big data, and edge computing. 
    
    Dr. Li is a recipient of the Stars of Tomorrow (Award of Excellent Intern), Microsoft Research Asia, Technology Talent of Xidian University, and AliStar.
\end{IEEEbiography}

\begin{IEEEbiography}[{\includegraphics[width=.95\linewidth]{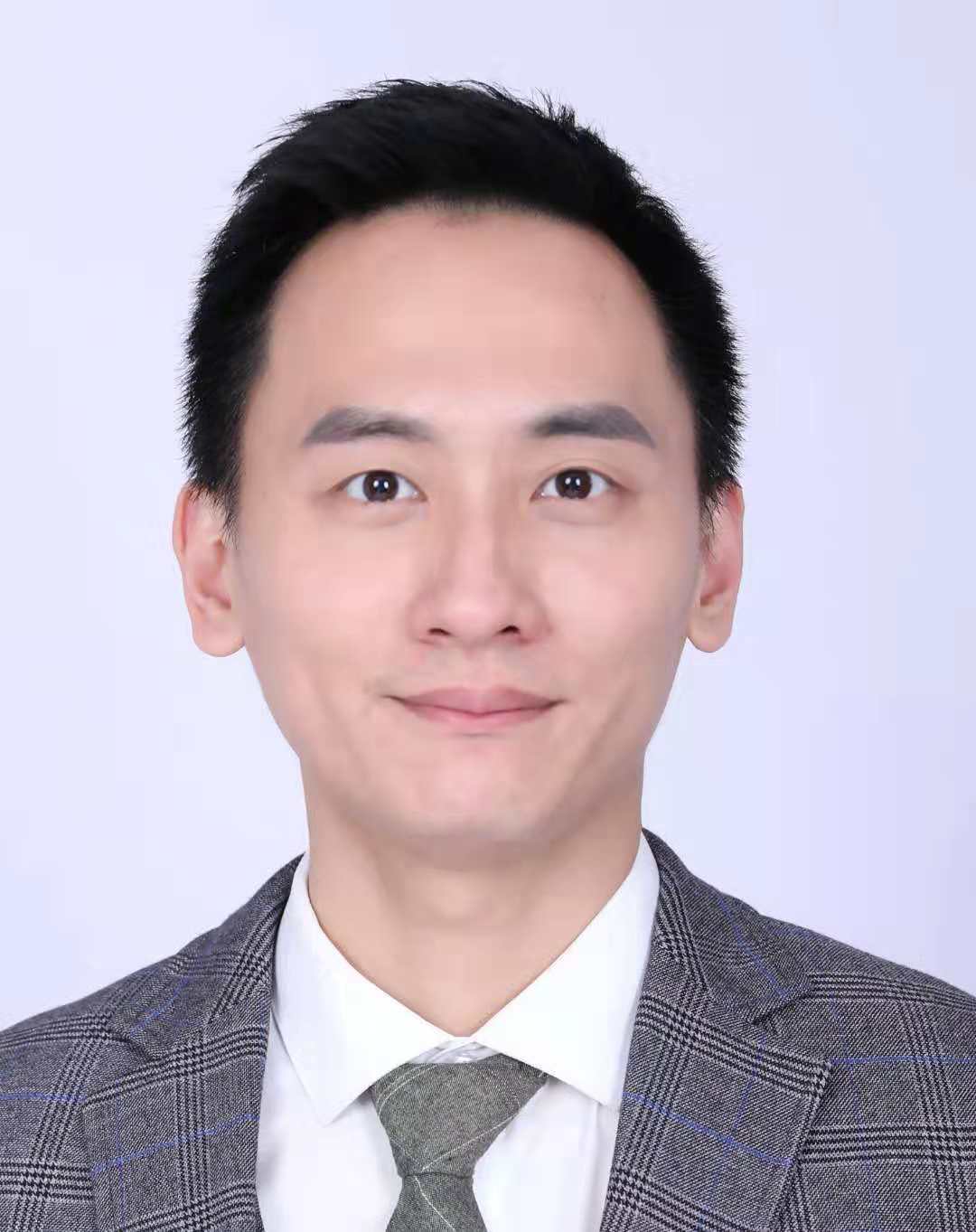}}]{Yifeng Zheng} (Member, IEEE)
    is an Assistant Professor with the School of Computer Science and Technology, Harbin Institute of Technology, Shenzhen, China. He received the PhD degree in computer science from the City University of Hong Kong, Hong Kong, in 2019. He worked as a postdoc with the Commonwealth Scientific and Industrial Research Organization (CSIRO), Australia and City University of Hong Kong. His work has appeared in prestigious venues such as ESORICS, DSN, ACM AsiaCCS, IEEE INFOCOM, IEEE ICDCS, IEEE Transactions on Dependable and Secure Computing, IEEE Transactions on Information Forensics and Security, and IEEE Transactions on Services Computing. He received the Best Paper Award in the European Symposium on Research in Computer Security (ESORICS) 2021. His current research interests are focused on security and privacy related to cloud computing, IoT, machine learning, and multimedia.
\end{IEEEbiography}

\begin{IEEEbiography}[{\includegraphics[width=.95\linewidth]{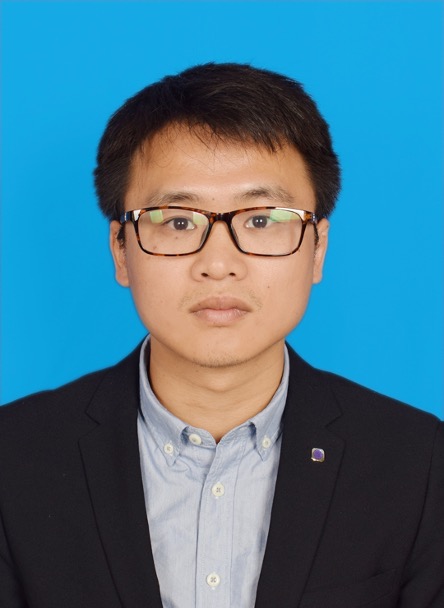}}]{Leo Yu Zhang} (M’17)
is currently a Senior Lecturer with the School of Information and Communication Technology, Griffith University, QLD, Australia. And he used to be a Lecturer at the School of Information Technology, Deakin University from 2018-2022. He received the bachelor’s and master’s degrees in computational mathematics from Xiangtan University, Xiangtan, China, in 2009 and 2012, respectively, and the Ph.D. degree from the City University of Hong Kong, Hong Kong, in 2016. He held various research positions with the City University of Hong Kong, the University of Macau, Macau, China, the University of Ferrara, Ferrara, Italy, and the University of Bologna, Bologna, Italy. His current research interests include trustworthy AI and applied cryptography, and he has published more than 90 refereed journal and conference articles in these fields.
\end{IEEEbiography}

\end{document}


\title{Supplementary Material}

\maketitle

\subsection{Trigger Sensitivity}

We design experiments to show this effect. Specifically, we generate new ``poisoned images'' with the residuals generated using other clean images and test their ASRs. For a clean image $x_i\in \bm{D}_{\textnormal{test}}$ and a trigger $t\in \bm{T}$, we randomly select image $x_j$, where $x_j\in \bm{D}_{\textnormal{test}}$ and  $x_j\neq x_i$, and feed $x_j$ and $t$ into the trigger embedding network $\mathcal{H}$ to generate a poisoned image $\tilde{x}_j$. Then, we construct a new ``poisoned image'' by adding the inconsistent residual between the poisoned image $\tilde{x}_j$ and its clean counterpart $x_j$ to the clean image $x_i$, i.e., $x_i+(\tilde{x}_j-x_j)$.
For simplicity, we set $j=(i+1)\bmod |\bm{D}_{\textnormal{test}}|$.

We use this manner to poison the three testing datasets and evaluate the average ASRs of the new ``poisoned images''. Tables~\ref{tab:appexclusiveness} and~\ref{tab:appexclusivenessimagenet} show that the ASRs decrease fast when the poisoned images are generated by adding inconsistent residuals. The results demonstrate that the residuals generated from our triggers are sample-specific and unique.

\begin{table}[!h]
\centering
\tiny
\caption{ASRs (\%) of the poisoned images on the MNIST, CIFAR-10 and GTSRB datasets with consistent and inconsistent residuals. }
\label{tab:appexclusiveness}
\begin{tabular}{lcccc} 
\toprule
Metrics        & MNIST & CIFAR-10 & GTSRB \\ 
\midrule
 1-to-4 (consistent) & 99.93 & 99.70& 98.89 \\
 1-to-4 (inconsistent) & 24.52 & 15.30 & 23.48\\
 10-to-4 (consistent) & 96.24 & 98.23 & 99.13 \\
 10-to-4 (inconsistent) &25.01 & 15.04& 7.84 \\
\bottomrule
\end{tabular}
\end{table}

\begin{table}[!h]
\centering
\footnotesize
\caption{\red{ASRs (\%) of the poisoned images on the ImageNet-10 dataset with consistent and inconsistent residuals. }}
\label{tab:appexclusivenessimagenet}
\begin{tabular}{lcccc} 
\toprule
\red{Metrics}     &    \red{ImageNet}\\ 
\midrule
 \red{2-to-4 (consistent)} &  \red{96.80}\\
 \red{2-to-4 (inconsistent)} & \red{26.20}\\
\bottomrule
\end{tabular}
\end{table}
 
\subsection{Fine-Pruning}

\begin{figure}[!t]
	\centering
	\begin{minipage}[b]{0.8\linewidth}
    	\begin{minipage}[b]{0.49\linewidth}
			\centerline{\includegraphics[width=1\linewidth]{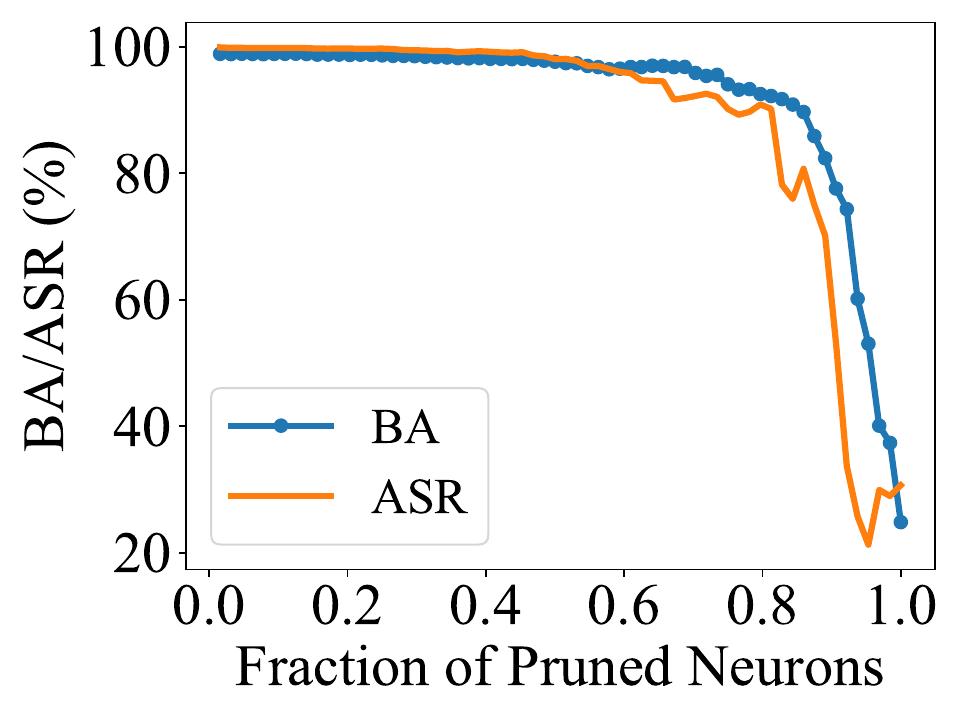}}
			\centerline{(a)}
		\end{minipage}	
            \begin{minipage}[b]{0.49\linewidth}
			\centerline{\includegraphics[width=1\linewidth]{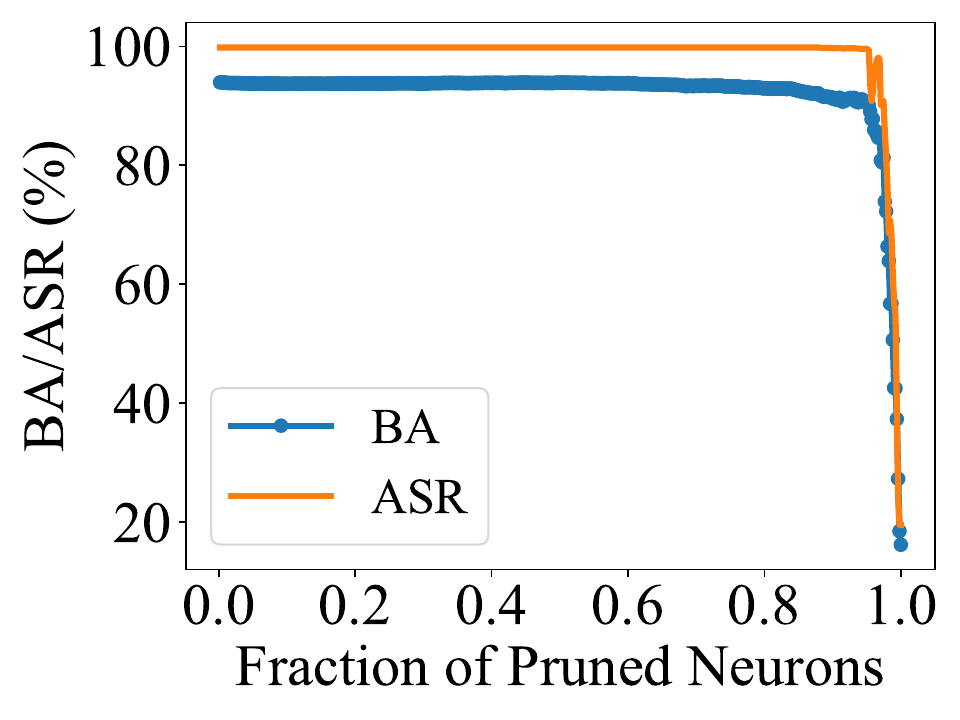}}
			\centerline{(b)}
		\end{minipage}
  \end{minipage}
    \begin{minipage}[b]{0.8\linewidth}
            \begin{minipage}[b]{0.49\linewidth}
			\centerline{\includegraphics[width=1\linewidth]{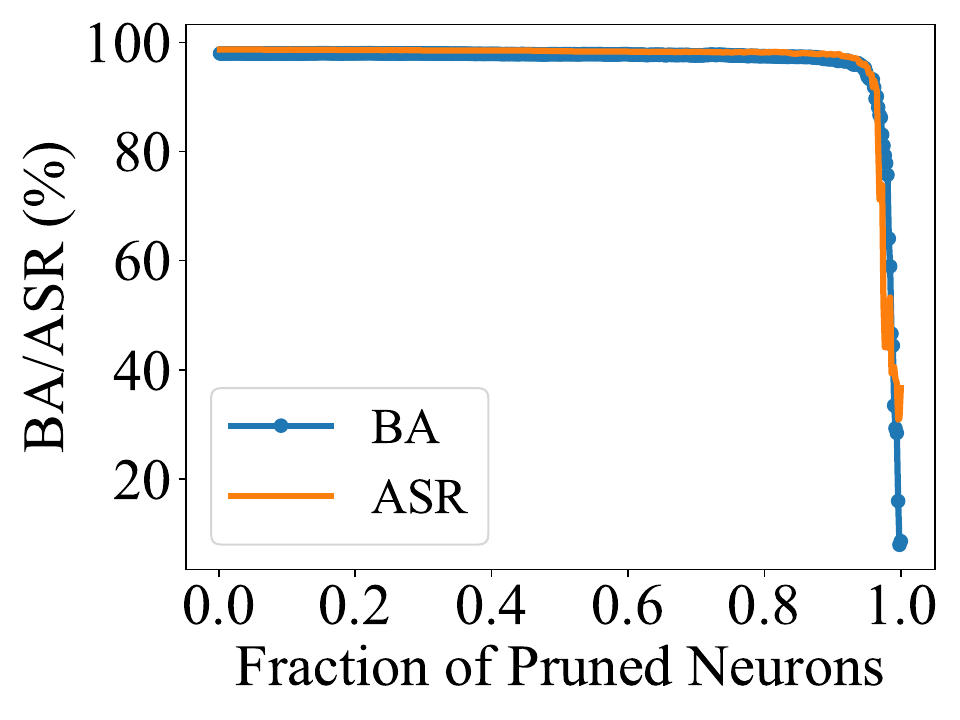}}
			\centerline{(c)}
		\end{minipage}
            \begin{minipage}[b]{0.49\linewidth}
			\centerline{\includegraphics[width=1\linewidth]{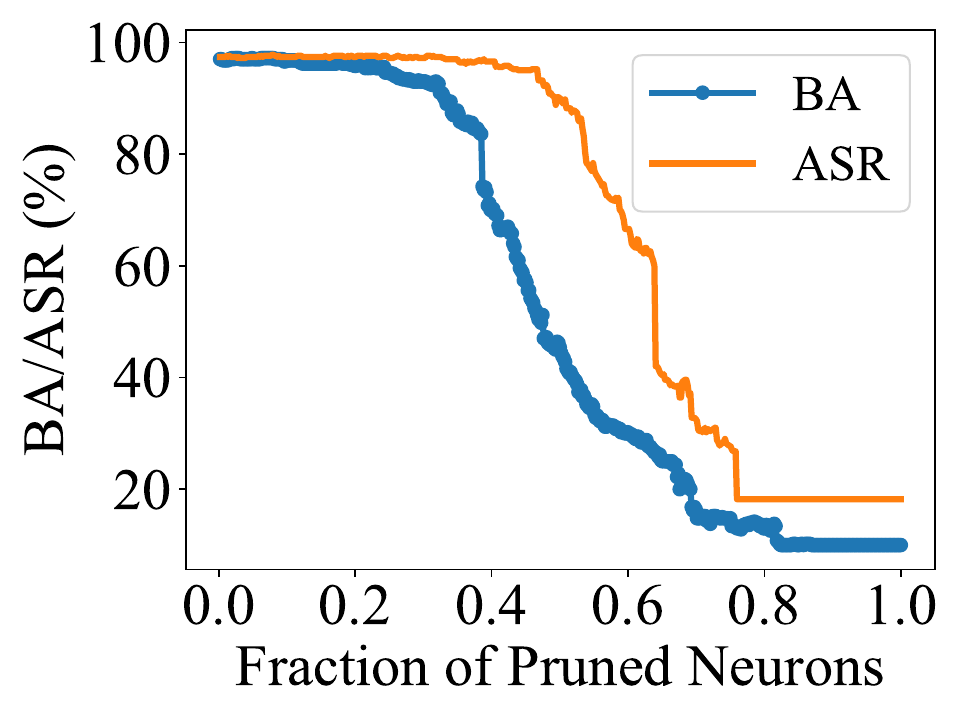}}
			\centerline{(d)}
		\end{minipage}
	\end{minipage}
	\caption{Ability of our $M$-to-$N$ backdoor attack to resist the fine-pruning defence on different datasets. Results with trigger number $M=10$ for (a) MNIST, (b) CIFAR-10, (c) GTSRB datasets, and $M=2$ for (d) ImageNet-10 dataset when simultaneously attacking five target classes.}
 \label{fig:fine-pruning}
\end{figure}
Fig.~\ref{fig:fine-pruning} displays the pruning effects of the $10$-to-$N$ attack when attacking five target classes simultaneously (i.e., $M=1,N=5$) on the four datasets. It can be seen that the ASRs of our attack only decrease slightly with the increase of the fraction of pruned neurons. 
Furthermore, the ASRs of our attack decrease after the BAs, indicating that FP cannot remove our backdoors without impacting the performance of original tasks.

\subsection{Neural Cleanse}
Fig.~\ref{fig: appncstrip}(a) shows that the average anomaly indices of our $10$-to-$N$ ($M=10$) attacks remain below the threshold of two, indicating that our attack is resistant against NC defense.
\subsection{STRIP}
\begin{figure}[h]
\centering
    \begin{minipage}[b]{0.48\linewidth}
			\centerline{\includegraphics[width=1\linewidth]{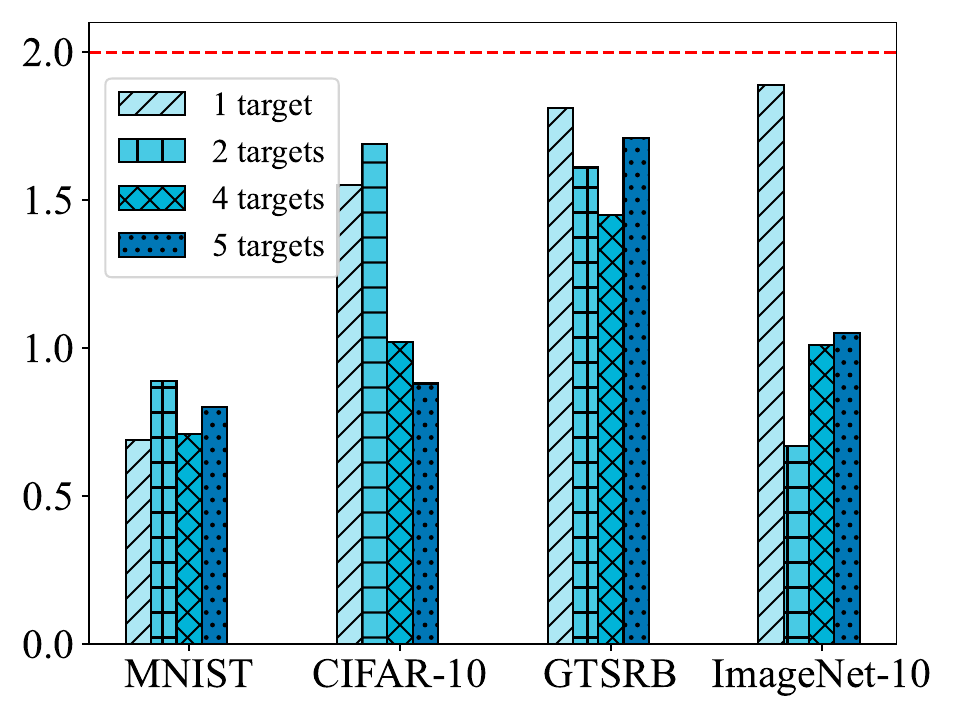}}
      \centerline{ (a)}
    \end{minipage}
 \begin{minipage}[b]{0.48\linewidth}
   \centerline{ \includegraphics[width=1\linewidth]{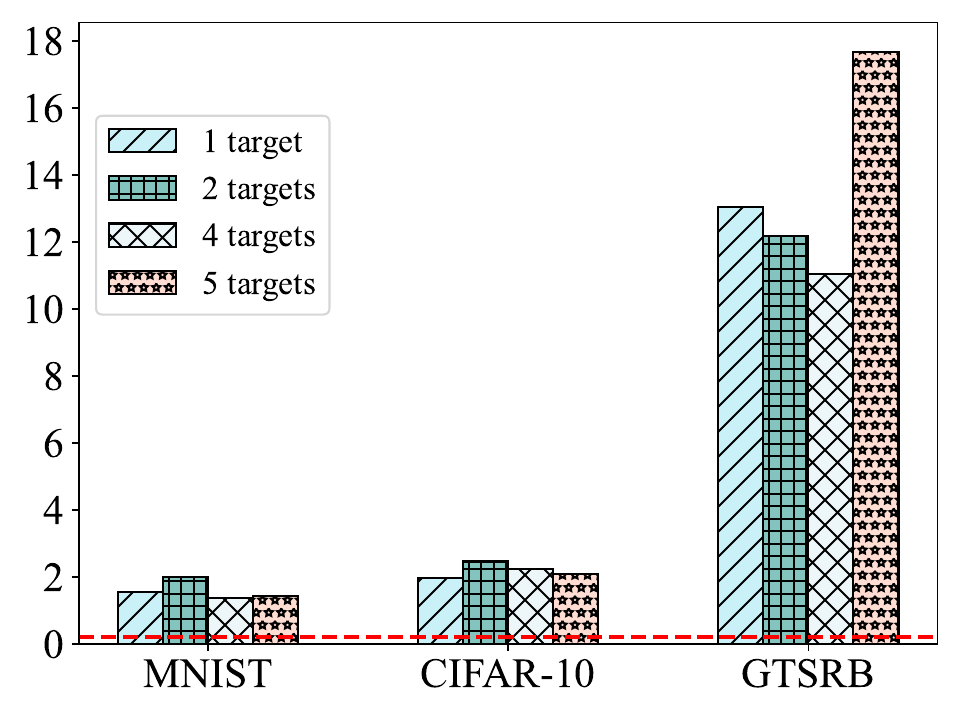}}
      \centerline{ (b)}
 \end{minipage}
  \caption{Ability of our $M$-to-$N$ backdoor attack to resist NC and STRIP. (a) The anomaly indexes of our attack with ten triggers ($M=10$) per target class. (b) The minimum entropy of our attack with ten triggers ($M=10$).}
  	\label{fig: appncstrip}
 \end{figure}

Fig.~\ref{fig: appncstrip}(b) shows that all entropy values of our $10$-to-$N$ ($M=10$) attacks exceed the detection threshold of 0.2, indicating that the STRIP identifies our poisoned images as clean images.

